\documentclass[3p, 10pt, onecolumn]{elsarticle}
\usepackage{graphicx,amscd,amsmath,amssymb,verbatim}
\usepackage{amsfonts,epsfig}
\usepackage{mathptmx}
\usepackage{setspace}
\usepackage{epsfig}
\usepackage{url}
\usepackage{multirow}
\usepackage{subfigure}
\usepackage{graphicx,color}
\usepackage[ruled,vlined]{algorithm2e}
\usepackage[normalem]{ulem}

\pdfoutput=1

\begin{document}

\journal{IJMPC}

\begin{frontmatter}

\title{Transfer matrix in counting problems}

\author{Roberto da Silva$^{1}$, Silvio R. Dahmen$^{1}$, J. R. Drugowich de
Fel\'{\i}cio$^{2}$}

\address{1 - Instituto de F\'{i}sica, Universidade Federal do Rio Grande do Sul, Porto Alegre, Rio Grande do Sul, Brazil\\ 
	2 - Departamento de F\'{i}sica, Faculdade de Filosofia, Ci\^{e}ncias e Letras de Riber\~{a}o Preto, Universidade de S\~{a}o Paulo, Ribeir\~{a}o Preto, S\~{a}o Paulo, Brazil}

\begin{abstract}

The transfer matrix is a powerful technique that can be applied to
statistical mechanics systems as, for example, in the calculus of the entropy
of the ice model. One interesting way to study such systems is to map it
onto a 3-color problem. In this paper, we explicitly build the transfer
matrix for the 3-color problem in order to calculate the number of possible
configurations for finite systems with free, periodic in one direction and
toroidal boundary conditions (periodic in both directions)

\end{abstract}


\begin{keyword}
 Transfer matrix \sep toroidal boundary conditions \sep ice-type model \sep three-color problem
\end{keyword}

\end{frontmatter}

\section{Introduction}

The transfer matrix technique in statistical physics was introduced by
Kramers and Wannier in 1941 in the context of two-dimensional ferromagnetic
systems \cite{Krammers-I,Krammers-II}. However its applicability extends
beyond spin models \cite{Dimarzio,Lieb-I,Lieb-II,Baxter,Pegg,Teif}. They are
very useful not only in the description of the thermodynamics of such
systems \cite{Nightingale,Batchelor}, but can also be applied to more
general settings, for example in Optics \cite{Katsidis2002} and Graph theory 
\cite{Engstrom2002}. Its foundation goes back to the roots of counting
problems in statistical mechanics. If in one dimension the application of
the concept in order to calculate the number of configurations is
straightforward, in higher dimensions the concept has its caveats. In this
paper, we wish to show how one may use the transfer matrix to study counting
problems of systems in lattices considering different boundary conditions.

\begin{figure}[tbp]
\begin{center}
\includegraphics[width=1.0\columnwidth]{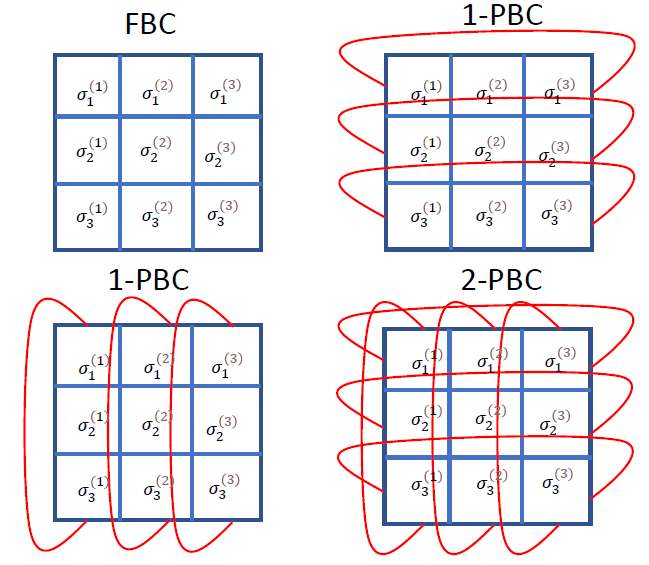}
\end{center}
\caption{Different boundary conditions that can be considered when using the
transfer matrix technique. }
\label{Fig:boundary_conditions}
\end{figure}

We can think of the problem in a more general way: we have an interacting
system where the objects (spins, or colors, or other objects that we can be
interested) are disposed in a two-dimensional rectangular lattice. For the
sake of simplicity, let us suppose a square lattice, $L\times L$. The number
of bonds between all pairs of nearest neighbor sites depends on the boundary
condition of the problem. If it is free, one has $2L(L-1)$ bonds, if
periodic in one of directions $L(2L-1)$, and if periodic in both directions,
one has $2L^{2}$ bonds, as exemplified in Fig. \ref{Fig:boundary_conditions}
for $L=3$.

Here, we are denoting $\sigma _{j}$ by $(\sigma _{1}^{(j)},...,\sigma
_{L}^{(j)})^{t}$ which is a particular configuration of $j$-th column. Now
comes the most important question of this article: how to compute the number
of possible configurations of the system for different boundary conditions.
The answer depends on what we have at hand and how the interactions between
the spins or other constituents are.

Let $T(\sigma _{i},\sigma _{j})$ denote the number of configurations
resulted from interaction between the column $\sigma _{i}$ with the column $%
\sigma _{j}$. $T$ is the so-called transfer matrix of the two-dimensional
system under analysis. If one has $L$ columns, represented respectively by
the vectors $\sigma _{1}$, $\sigma _{2}$, ..., $\sigma _{L}$, and the
interaction takes place only between nearest neighbors, the number of
possible configurations is given by:%
\begin{equation}
\Omega(L)=\sum_{\sigma _{1}}\sum_{\sigma _{2}}...\sum_{\sigma _{L}}T(\sigma
_{1},\sigma _{2})T(\sigma _{2},\sigma _{3})....T(\sigma _{L-1},\sigma
_{L})T(\sigma _{1},\sigma _{L})  \label{Eq:Transfer_matrix_first}
\end{equation}%
where the sum is performed over all possible values of $\sigma _{1}$, ...,$%
\sigma _{L}$.

This result certainly depends on how $T(\sigma _{i},\sigma _{j})$ was built,
i.e. whether we consider free or periodic boundary conditions. If we
consider free boundary conditions, that is all interactions between
successive elements along the column are accounted for except between $%
\sigma_{L}^{(j)}$ and\ $\sigma _{1}^{(j)}$, we denote the transfer matrix by 
$T_{FBC}$. On the other hand, if the interaction between $\sigma _{L}^{(j)}$
and\ $\sigma _{1}^{(j)}$ is also included, thus we will denote the transfer
matrix by $T_{PBC}$.

The question now is how to obtain the different cases depicted in Fig. \ref%
{Fig:boundary_conditions} from $T_{FBC}$ and $T_{PBC}$. Actually, there are
only three different cases since the situations of Fig. \ref%
{Fig:boundary_conditions} (b) and Fig. \ref{Fig:boundary_conditions} (c) are
symmetric. For that, we will simply use the term FBC to describe the case of
Fig. \ref{Fig:boundary_conditions} (a), 1-PBC will be used to describe the
situations that present periodic boundary condition in only one direction:
Fig \ref{Fig:boundary_conditions} (b) and Fig \ref{Fig:boundary_conditions}
(c), and finally 2-PBC describes the situation of the Fig. \ref%
{Fig:boundary_conditions} (d), i.e., periodic boundary conditions in both
directions.

For example, from $T_{FBC}$ we can obtain the counting $\Omega$ for FBC or
1-PBC and from $T_{PBC}\ $we can obtain $\Omega$ for 1-PBC or 2-PBC. If we
want PBC in the second direction (along the rows) we simply attribute the
value $T(\sigma _{1},\sigma _{L})=1$.

Our first case is to calculate $\Omega_{FBC}(L)$, the number of
configurations with FBC. Thus using the multiplicative principle and
considering the sum over all configurations and the considerations above,
one has:%
\begin{equation}
\begin{array}{lll}
\Omega_{FBC}(L) & = & \sum_{\sigma _{1}}\sum_{\sigma _{2}}...\sum_{\sigma
_{L}}T_{FBC}(\sigma _{1},\sigma _{2})T_{FBC}(\sigma _{2},\sigma
_{3})....T_{FBC}(\sigma _{L-1},\sigma _{L}) \\ 
&  &  \\ 
& = & \sum_{\sigma _{1}}\sum_{\sigma _{L}}T_{FBC}^{L-1}(\sigma _{1},\sigma
_{2}) \\ 
&  &  \\ 
& = & Add(T_{FBC}^{L-1})\text{,}%
\end{array}
\label{Eq:FBC}
\end{equation}%
where $Add(X)$ denotes the sum of all elements of the matrix $X.$

Naturally, $\Omega_{1-PBC}(L)$ can be also calculated using $T_{FBC}$. In this
case $T_{FBC}(\sigma _{L},\sigma _{1})$ is not set to 1 in the expression.
From this we obtain the important cyclical property of the trace: 
\begin{equation}
\begin{array}{lll}
\Omega_{1-PBC}(L) & = & \sum_{\sigma _{1}}\sum_{\sigma _{2}}...\sum_{\sigma
_{L}}T_{FBC}(\sigma _{1},\sigma _{2})T_{FBC}(\sigma _{2},\sigma
_{3})....T_{FBC}(\sigma _{L-1},\sigma _{L})T_{FBC}(\sigma _{L},\sigma _{1})
\\ 
&  &  \\ 
& = & Tr(T_{FBC}^{L})%
\end{array}
\label{Eq:1-PBC-trace}
\end{equation}

However, if we already have periodic boundary conditions in one direction in
($T_{PBC}$) we can also obtain $\Omega_{1-PBC}(L)$. However in this case we take $%
T_{PBC}(\sigma _{1},\sigma _{L})=1$:

\begin{equation}
\begin{array}{lll}
\Omega_{1-PBC}(L) & = & \sum_{\sigma _{1}}\sum_{\sigma _{2}}...\sum_{\sigma
_{L}}T_{PBC}(\sigma _{1},\sigma _{2})T_{PBC}(\sigma _{2},\sigma
_{3})....T_{PBC}(\sigma _{L-1},\sigma _{L}) \\ 
&  &  \\ 
& = & \sum_{\sigma _{1}}\sum_{\sigma _{L}}T_{PBC}^{L-1}(\sigma _{1},\sigma
_{2}) \\ 
&  &  \\ 
& = & Add(T_{PBC}^{L-1}).%
\end{array}
\label{Eq:1-PBC-without-trace}
\end{equation}%
With this we can calculate $\Omega_{1-PBC}(L)$ in two different ways: using $%
T_{FBC}$ (Eq. \ref{Eq:1-PBC-trace}) or $T_{PBC}$ (Eq. \ref%
{Eq:1-PBC-without-trace}). Finally to obtain $\Omega_{2-PBC}(L)$ one has only one
possibility: one starts with $T_{PBC}$ and completes it to obtain the PBC in
the other direction:

\begin{equation}
\begin{array}{lll}
\Omega_{2-PBC}(L) & = & \sum_{\sigma _{1}}\sum_{\sigma _{2}}...\sum_{\sigma
_{L}}T_{PBC}(\sigma _{1},\sigma _{2})T_{PBC}(\sigma _{2},\sigma
_{3})....T_{PBC}(\sigma _{L-1},\sigma _{L})T_{PBC}(\sigma _{L},\sigma _{1})
\\ 
&  &  \\ 
& = & Tr(T_{FBC}^{L})%
\end{array}
\label{Eq:2-PBC}
\end{equation}

In a recent and more didactic work \cite{rdasilvaEJP}, we showed how to
implicitly use the transfer matrix method to obtain the entropy of the
two-dimensional ice-type model mapping the problem onto the three-color
problem. However two important points were not considered in our approach:

\begin{enumerate}
\item The method did not explicitly explore the properties of the matrices,
something which is done in detail in the present work;

\item Moreover, we present the results for the case of toroidal boundary
conditions, extending Creswick's method \cite{Creswick} that considers
periodic boundary conditions in one direction only.
\end{enumerate}

We obtain $\Omega(L)$ for different boundary conditions only by switching between
equations \ref{Eq:FBC}, \ref{Eq:1-PBC-trace}, \ref{Eq:1-PBC-without-trace},
and \ref{Eq:2-PBC}. We also show that using periodic boundary conditions in
both directions yields better estimates than previously obtained in \cite%
{rdasilvaEJP}. The paper is organized as follows: for the sake of
completeness we present in the next section the ice-type model, a
two-dimensional structure proposed to explain the residual entropy of ice at 
$T=0$~\cite{Pauling}. Finally, we show how this problem can be mapped in the
three-color problem.

We then present the simplest cases ($L=1$, $2$, and $3$) and we explicitly
evaluate $\Omega (L)$ for each case. In what follows, we present the results
considering an efficient computer routine for cases with $L>3$ and we
estimate the entropy of the system extrapolating $L\rightarrow \infty $ for
every boundary condition considered. Finally, we present some considerations
and conclusions suggesting that our approach can be extended to other
counting problems in Statistical Physics.

\section{The ice-type model and its mapping onto the three-color problem}

\begin{figure}[tbp]
\begin{center}
\includegraphics[width=1.0%
\columnwidth]{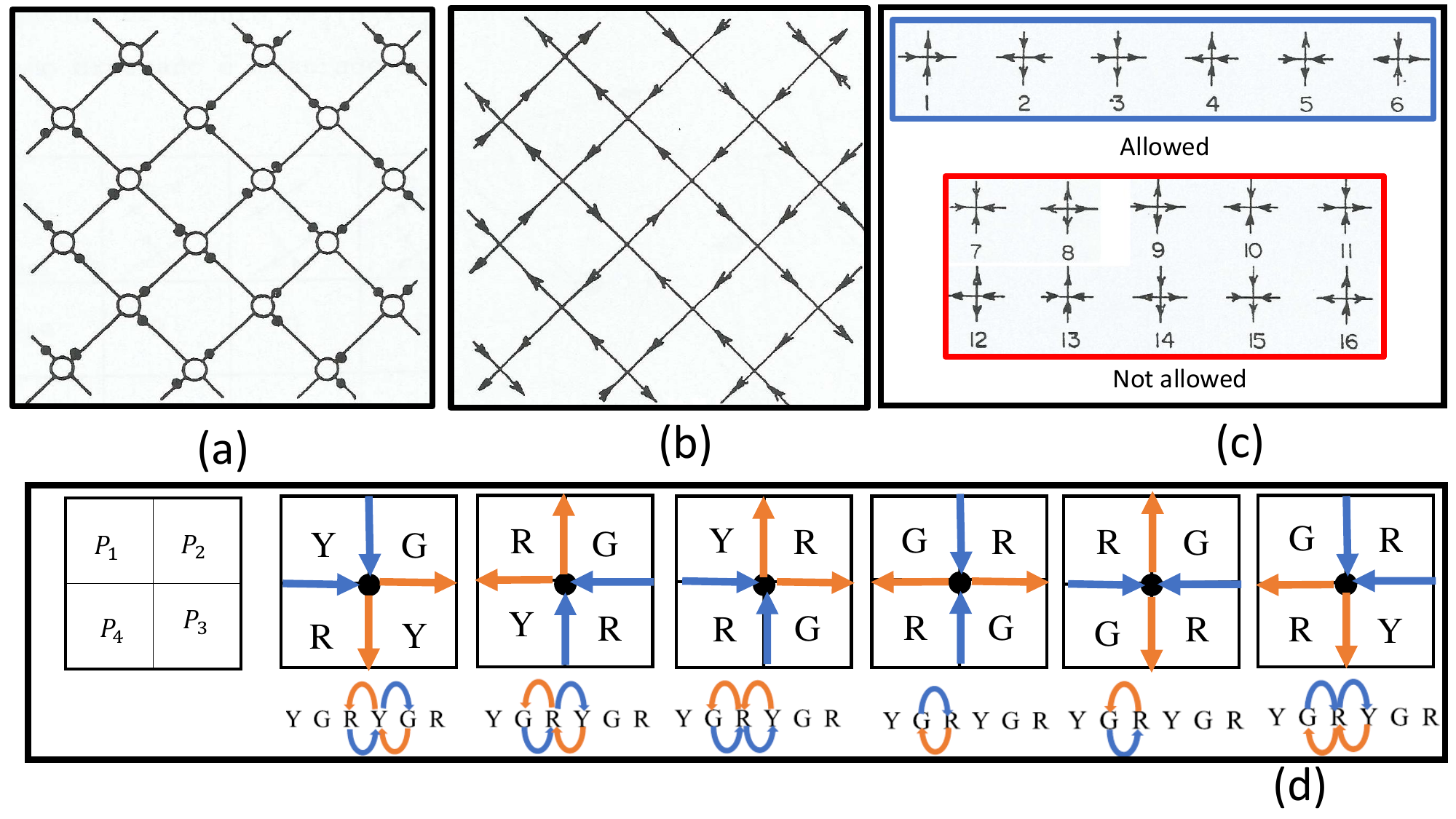}
\end{center}
\caption{(a) The two-dimensional ice model. White circles represent oxygen
atoms while black ones stand for hydrogen; (b) Representation of (a) using
arrows; (c) the six different configurations of arrows allowed in the
ice-model and the prohibited ones; (d) mapping of the six-vertex model onto
the three-color problem.}
\label{Fig:mapeamento_gelo-seis_vertices-cores}
\end{figure}

Ice has a tetrahedral structure with oxygen atoms occupying the vertices.
Each oxygen is linked by hydrogen bonds to four other oxygen atoms. This
means that if we wish to consider a two-dimensional version of this model it
has to preserve the fundamental characteristics of the real structure which
is the fact that each oxygen atom has four neighbours. Since each water
molecule has only two hydrogen atoms, in the real structure two of these
hydrogen atoms must be in the nearest equilibrium position and the two other
at the larger distance. In Fig. \ref{Fig:mapeamento_gelo-seis_vertices-cores}
(a) we show a two dimensional representation of this ice-type model. White
circles represent $O$ atoms, while black ones represent hydrogen ones.

Since each hydrogen atom can be in two distinct positions, Pauling \cite%
{Pauling} introduced an arrow to indicate whether it is near to (incoming
arrow) or far from (outcoming arrow) an oxygen. The percentage of H$_{3}$O$%
^{+}$ and OH$^{-}$ ions are taken as zero which means that each $O$ atom
(site in the lattice) must necessarily have 2 and only 2 hydrogen atoms next
to it (two incoming and two outcoming arrows from a vertex). Fig. \ref%
{Fig:mapeamento_gelo-seis_vertices-cores} (b) is the same as Fig. \ref%
{Fig:mapeamento_gelo-seis_vertices-cores} (a) represented as arrows. From
the 16 possible types of vertices, only six satisfy the so-called `ice
rules' (Fig. \ref{Fig:mapeamento_gelo-seis_vertices-cores} (c)), introduced
by Bernal and Fowler in 1933 \cite{Bernal} and improved by Linus Pauling 
\cite{Pauling}.

On the other hand, it is very interesting to recast this problem using the
language of map coloring. For that, we attribute three colors, yellow ($Y$),
green($G$), and red($R$) to countries in a world represented by a two
dimensional lattice. For example a 2$\times$2--world with $4$ countries is
represented in Fig. \ref{Fig:mapeamento_gelo-seis_vertices-cores} (d), where 
$P_{1}$, $P_{2}$, $P_{3}$, and $P_{4}$ denote countries. Let us call a 
\textit{proper} coloring of the map one where countries with a common border
are colored differently while in the opposite case they could have the same
color. For example $P_{1}$ must be different from $P_{2}$ and $P_{4}$, but
it could have the same color as $P_{3}$. In the same Fig. \ref%
{Fig:mapeamento_gelo-seis_vertices-cores} (d) we can observe 6 particular
examples of proper colorings of a 2 $\times $ 2--world with a maximal of
three colors, resulting in the six possible different arrow configurations
of the six-vertex model.

We shall use a cyclical convention for the colours: $Y$ follows $G$, $G$
follows $R$, and $R$ follows $Y$ ($YGRYGRYGR$...). Every time we rotate the
map clockwise with respect to a perpendicular axis going through the common
vertex (black dot) of each 2 $\times $ 2--map inserted in the whole L$\times$%
L--map. Starting for example from $P_{1}$, if we change from yellow to green
(green to red, red to yellow) the arrow (depicted blue) in the boundary will
be directed to the common point, while the changing from red to green (green
to yellow, yellow to red) is represented by an orange arrow leaving the
common point. This applies both for the horizontal as well as the vertical
boundaries. This common point that resembles the position of oxygen atoms
will always have two arrows in and two arrows out. Fig. \ref%
{Fig:mapeamento_gelo-seis_vertices-cores} (d) shows the six possible
configurations of arrows and colors on a map. Let us consider the first
color configuration in that same figure. We start in $P_{1}$ with $Y$ and $%
P_{1}$ with $G$. Going clockwise there is a blue vertical arrow pointing to
the vertex. From $P_{2}$ to $P_{3}$, one has $G$ to $Y$ in the
counter-clockwise direction, and thus there is an orange horizontal arrow
off the vertex. From $P_{3}$ to $P_{4}$ there is an orange vertical arrow
off the vertex, and from $P_{4}$ to $P_{1}$ there is a blue horizontal
incoming arrow. Other colorings (there are another 12 different colorings in
addition to the six shown in the example) in this mapping only lead to one
of the possible arrow configurations depicted in this same figure.

In fact, each configuration of arrows corresponds to three possible
colorings in fig. \ref{Fig:mapeamento_gelo-seis_vertices-cores} (d), since
there are three possible colors for P1 to start with. Thus we can write

\begin{equation}
\Omega _{\text{colours}}(L)=3\Omega _{\text{six-vertex}}(L).
\end{equation}

Denoting $L^{2}=N$, we define $\Omega _{\text{six-vertex}}(L)=W^{N}$ and
thus the entropy of the ice-type model is given by: 
\begin{equation}
S=k_{B}\ln \Omega _{\text{six-vertex}}=N\,k_{B}\ln W
\end{equation}

The idea now is to estimate $W$ by calculating $\Omega_{\text{colours}}(L)$
through an extrapolation via the transfer matrix. But before doing that, we
can better understand how to estimate $\Omega_{\text{colours}}(L)$ if we
look at some simple cases explicitly.

\section{Some explicit examples: Pedagogical aspects}

Let us show how the method works by constructing and calculating $T_{FBC}$
and $T_{PBC}$ for the simplest non trivial cases $L=2$ and $L=3$ and after.
For $L=2$, note that $T_{FBC}=T_{PBC}$. We thus use simply $T$ for the
transfer matrix since we compose it by crossing states $\left\vert \varphi
\right\rangle =\left\vert C_{1}C_{2}\right\rangle $. For a proper coloring, $%
C_{1}$ and $C_{2}$ must be different.

With 3 colors, of a total of 9 possible configurations only $n_{\max }=6$
different states are allowed. For example $\left\vert RG\right\rangle $ is a
possible state while $\left\vert RR\right\rangle $ is not. Thus for two
states $\left\vert \varphi _{1}\right\rangle =\left\vert
C_{1}^{(1)}C_{2}^{(1)}\right\rangle $ and $\left\vert \varphi
_{2}\right\rangle =\left\vert C_{1}^{(2)}C_{2}^{(2)}\right\rangle $, the
scalar product $\left\langle \varphi _{1}|\varphi _{2}\right\rangle
=\left\langle \varphi _{2}|\varphi _{1}\right\rangle $ is defined by: 
\begin{equation}
\left\langle \varphi _{1}|\varphi _{2}\right\rangle =\left\{ 
\begin{array}{ll}
1 & \text{if\ }C_{1}^{(1)}\neq C_{1}^{(2)}\text{ and }C_{2}^{(1)}\neq
C_{2}^{(2)} \\ 
&  \\ 
0 & \text{otherwise}%
\end{array}%
\right.
\end{equation}

We can thus explicitly write $T(L=2)$ as:

\begin{equation}
T=\left( 
\begin{array}{ccccccc}
& \left\vert YR\right\rangle & \left\vert RY\right\rangle & \left\vert
GR\right\rangle & \left\vert RG\right\rangle & \left\vert YG\right\rangle & 
\left\vert GY\right\rangle \\ 
\left\langle YR\right\vert & 0 & 1 & 0 & 1 & 0 & 1 \\ 
\left\langle RY\right\vert & 1 & 0 & 1 & 0 & 1 & 0 \\ 
\left\langle GR\right\vert & 0 & 1 & 0 & 1 & 1 & 0 \\ 
\left\langle RG\right\vert & 1 & 0 & 1 & 0 & 0 & 1 \\ 
\left\langle YG\right\vert & 0 & 1 & 1 & 0 & 0 & 1 \\ 
\left\langle GY\right\vert & 1 & 0 & 0 & 1 & 1 & 0%
\end{array}%
\right)  \label{Eq:Matrix_L=2}
\end{equation}
\noindent In this particular case $C_{FBC}=C_{1-PBC}=C_{2-PBC}$, and
therefore, $Tr(T^{2})=Add(T)=18$. Is it easy to understand this result: in a
world with 4 countries ($L=2$) and 3 colors we can paint $P_{1}$ and $P_{3}$
with same or with different colors. If they have the same color, there are 3
colors to choose and $P_{2}$ and $P_{4}$, which of course must be different
from the previous ones. The number of ways is thus: 
\begin{equation}
f_{=}(3)=\underset{P_{1}\text{ and }P_{3}}{\underbrace{3}}\times \underset{%
P_{2}\text{ }}{\underbrace{2}}\times \underset{P_{4}\text{ }}{\underbrace{2}}%
=12
\end{equation}

The other possibility is to put different colors in $P_{1}$ and $P_{3}$. In
this case $P_{2}$ and $P_{4}$, which must differ from the former, can only
be painted in one color. Thus: 
\begin{equation}
f_{\neq }(3)=\underset{P_{1}\text{ and }P_{3}}{\underbrace{3\times 2}}\times 
\underset{P_{2}\text{ }}{\underbrace{1}}\times \underset{P_{4}\text{ }}{%
\underbrace{1}}=6
\end{equation}

Consequently the number of different colorings oa f $2\times2$--world is 
\begin{equation}
f(3)=f_{=}(3)+f_{\neq }(3)=18,
\end{equation}%
which is exactly the number we found via matrix operations.

One may also observe that such result can be checked in two other ways, if
our prescriptions are correct: a) computing $T^{2}(L=2)$ and directly taking
the trace of $T^{2}$, or, b) calculating the eigenvalues of $T$, and using
the fact that $Tr(T^{n})=\sum_{i=1}^{n_{\max }}\lambda _{i}^{n}$.
Calculating $T^{2}$ explicitly, one has 
\begin{equation}
T^{2}=\left(\allowbreak 
\begin{array}{cccccc}
3 & 0 & 2 & 1 & 2 & 1 \\ 
0 & 3 & 1 & 2 & 1 & 2 \\ 
2 & 1 & 3 & 0 & 1 & 2 \\ 
1 & 2 & 0 & 3 & 2 & 1 \\ 
2 & 1 & 1 & 2 & 3 & 0 \\ 
1 & 2 & 2 & 1 & 0 & 3%
\end{array}%
\right) \allowbreak
\end{equation}%
One may check that $Tr(T^{2})=6\times 3=18$, which matches with our previous
results. Calculating the eigenvalues of $T$ one has: $\lambda _{1}=\lambda
_{2}=0$, $\lambda _{3}=3$, $\lambda _{4}=1$, and $\lambda _{5}=\lambda
_{6}=-2$, and $0^{2}+0^{2}+3^{2}+1^{2}+(-2)^{2}+(-2)^{2}=18$ as expect. The
case $L=2$ is simple and more investment is necessary to understand the
method.

To better understand the procedure and check the conjectures, we must study
the case $L=3$ where we have $T_{FBC}\neq T_{PBC}$. Defining now states $%
\left\vert \varphi \right\rangle =\left\vert C_{1}C_{2}C_{3}\right\rangle $,
we have a total of 27 candidate states. Here we can consider two
possibilities: $C_{1}$ must be different from $C_{3}$ (PBC) or not (FBC). In
the case of $T_{FBC}$, we have $n_{\max }=12$ different possible states $%
\left\vert C_{1}C_{2}C_{3}\right\rangle $ since the color $C_{1}$ can be
equal to $C_{3} $, and the only restriction is in the situations $C_{1}\neq
C_{2}$ and $C_{2}\neq C_{3}$. Thus we have:

\begin{equation}
T_{FBC}=\left( 
\begin{array}{ccccccccccccc}
\text{{\tiny {}}} & \text{{\tiny $\left\vert GYR\right\rangle $}} & \text{%
{\tiny $\left\vert RYG\right\rangle $}} & \text{{\tiny $\left\vert
RYR\right\rangle $}} & \text{{\tiny $\left\vert GYG\right\rangle $}} & \text{%
{\tiny $\left\vert GRY\right\rangle $}} & \text{{\tiny $\left\vert
YRG\right\rangle $}} & \text{{\tiny $\left\vert GRG\right\rangle $}} & \text{%
{\tiny $\left\vert YRY\right\rangle $}} & \text{{\tiny $\left\vert
YGR\right\rangle $}} & \text{{\tiny $\left\vert RGY\right\rangle $}} & \text{%
{\tiny $\left\vert YGY\right\rangle $}} & \text{{\tiny $\left\vert
RGR\right\rangle $}} \\ 
\text{{\tiny $\left\langle GYR\right\vert $}} & {\tiny 0} & {\tiny 0} & 
{\tiny 0} & {\tiny 0} & {\tiny 0} & {\tiny 1} & {\tiny 0} & {\tiny 1} & 
{\tiny 0} & {\tiny 1} & {\tiny 1} & {\tiny 0} \\ 
\text{{\tiny $\left\langle RYG\right\vert $}} & {\tiny 0} & {\tiny 0} & 
{\tiny 0} & {\tiny 0} & {\tiny 1} & {\tiny 0} & {\tiny 0} & {\tiny 1} & 
{\tiny 1} & {\tiny 0} & {\tiny 1} & {\tiny 0} \\ 
\text{{\tiny $\left\langle RYR\right\vert $}} & {\tiny 0} & {\tiny 0} & 
{\tiny 0} & {\tiny 0} & {\tiny 1} & {\tiny 1} & {\tiny 1} & {\tiny 1} & 
{\tiny 0} & {\tiny 0} & {\tiny 1} & {\tiny 0} \\ 
\text{{\tiny $\left\langle GYG\right\vert $}} & {\tiny 0} & {\tiny 0} & 
{\tiny 0} & {\tiny 0} & {\tiny 0} & {\tiny 0} & {\tiny 0} & {\tiny 1} & 
{\tiny 1} & {\tiny 1} & {\tiny 1} & {\tiny 1} \\ 
\text{{\tiny $\left\langle GRY\right\vert $}} & {\tiny 0} & {\tiny 1} & 
{\tiny 1} & {\tiny 0} & {\tiny 0} & {\tiny 0} & {\tiny 0} & {\tiny 0} & 
{\tiny 1} & {\tiny 1} & {\tiny 0} & {\tiny 1} \\ 
\text{{\tiny $\left\langle YRG\right\vert $}} & {\tiny 1} & {\tiny 0} & 
{\tiny 1} & {\tiny 0} & {\tiny 0} & {\tiny 0} & {\tiny 0} & {\tiny 0} & 
{\tiny 0} & {\tiny 1} & {\tiny 0} & {\tiny 1} \\ 
\text{{\tiny $\left\langle GRG\right\vert $}} & {\tiny 0} & {\tiny 0} & 
{\tiny 1} & {\tiny 0} & {\tiny 0} & {\tiny 0} & {\tiny 0} & {\tiny 0} & 
{\tiny 1} & {\tiny 1} & {\tiny 1} & {\tiny 1} \\ 
\text{{\tiny $\left\langle YRY\right\vert $}} & {\tiny 1} & {\tiny 1} & 
{\tiny 1} & {\tiny 1} & {\tiny 0} & {\tiny 0} & {\tiny 0} & {\tiny 0} & 
{\tiny 0} & {\tiny 0} & {\tiny 0} & {\tiny 1} \\ 
\text{{\tiny $\left\langle YGR\right\vert $}} & {\tiny 0} & {\tiny 1} & 
{\tiny 0} & {\tiny 1} & {\tiny 1} & {\tiny 0} & {\tiny 1} & {\tiny 0} & 
{\tiny 0} & {\tiny 0} & {\tiny 0} & {\tiny 0} \\ 
\text{{\tiny $\left\langle RGY\right\vert $}} & {\tiny 1} & {\tiny 0} & 
{\tiny 0} & {\tiny 1} & {\tiny 1} & {\tiny 1} & {\tiny 1} & {\tiny 0} & 
{\tiny 0} & {\tiny 0} & {\tiny 0} & {\tiny 0} \\ 
\text{{\tiny $\left\langle YGY\right\vert $}} & {\tiny 1} & {\tiny 1} & 
{\tiny 1} & {\tiny 1} & {\tiny 0} & {\tiny 0} & {\tiny 1} & {\tiny 0} & 
{\tiny 0} & {\tiny 0} & {\tiny 0} & {\tiny 0} \\ 
\text{{\tiny $\left\langle RGR\right\vert $}} & {\tiny 0} & {\tiny 0} & 
{\tiny 0} & {\tiny 1} & {\tiny 1} & {\tiny 1} & {\tiny 1} & {\tiny 1} & 
{\tiny 0} & {\tiny 0} & {\tiny 0} & {\tiny 0}%
\end{array}%
\right)  \label{Eq:FBC-3}
\end{equation}

Calculating $T_{FBC}^{2}$, one has: 
\begin{equation}
T_{FBC}^{2}=\left( \allowbreak 
\begin{array}{cccccccccccc}
4 & 2 & 3 & 3 & 1 & 1 & 2 & 0 & 0 & 1 & 0 & 2 \\ 
2 & 4 & 3 & 3 & 1 & 0 & 2 & 0 & 1 & 1 & 0 & 2 \\ 
3 & 3 & 5 & 2 & 0 & 0 & 1 & 0 & 2 & 3 & 1 & 4 \\ 
3 & 3 & 2 & 5 & 3 & 2 & 4 & 1 & 0 & 0 & 0 & 1 \\ 
1 & 1 & 0 & 3 & 5 & 3 & 4 & 3 & 1 & 0 & 2 & 0 \\ 
1 & 0 & 0 & 2 & 3 & 4 & 3 & 3 & 0 & 1 & 2 & 0 \\ 
2 & 2 & 1 & 4 & 4 & 3 & 5 & 2 & 0 & 0 & 1 & 0 \\ 
0 & 0 & 0 & 1 & 3 & 3 & 2 & 5 & 2 & 2 & 4 & 1 \\ 
0 & 1 & 2 & 0 & 1 & 0 & 0 & 2 & 4 & 3 & 3 & 3 \\ 
1 & 1 & 3 & 0 & 0 & 1 & 0 & 2 & 3 & 5 & 3 & 4 \\ 
0 & 0 & 1 & 0 & 2 & 2 & 1 & 4 & 3 & 3 & 5 & 2 \\ 
2 & 2 & 4 & 1 & 0 & 0 & 0 & 1 & 3 & 4 & 2 & 5%
\end{array}%
\right)
\end{equation}

So if one calculates $Add(T_{FBC}^{2})$, it yields the number of
configurations of a $3\times3$--map with free boundary conditions in both
directions. This results in $246$. This can be checked by computing all
possibilities with a simple algorithm (see for example \cite{rdasilvaEJP}).
We will check this when we find $T_{PBC}$. However if one wants to obtain
the result with periodic boundary conditions in one direction, the only
possibility is to take $Tr(T_{FBC}^{3})$ . Evaluating $T_{FBC}^{3}$
explicitly, one has: 
\begin{equation}
T_{FBC}^{3}=\left( \allowbreak 
\begin{array}{cccccccccccc}
2 & 1 & 4 & 3 & 8 & 10 & 6 & 14 & 8 & 11 & 14 & 7 \\ 
1 & 2 & 3 & 4 & 11 & 8 & 7 & 14 & 10 & 8 & 14 & 6 \\ 
4 & 3 & 2 & 10 & 17 & 15 & 15 & 17 & 6 & 6 & 14 & 3 \\ 
3 & 4 & 10 & 2 & 6 & 6 & 3 & 14 & 15 & 17 & 17 & 15 \\ 
8 & 11 & 17 & 6 & 2 & 1 & 3 & 5 & 13 & 16 & 9 & 18 \\ 
10 & 8 & 15 & 6 & 1 & 2 & 3 & 3 & 8 & 13 & 6 & 15 \\ 
6 & 7 & 15 & 3 & 3 & 3 & 2 & 9 & 15 & 18 & 14 & 18 \\ 
14 & 14 & 17 & 14 & 5 & 3 & 9 & 2 & 6 & 9 & 3 & 14 \\ 
8 & 10 & 6 & 15 & 13 & 8 & 15 & 6 & 2 & 1 & 3 & 3 \\ 
11 & 8 & 6 & 17 & 16 & 13 & 18 & 9 & 1 & 2 & 5 & 3 \\ 
14 & 14 & 14 & 17 & 9 & 6 & 14 & 3 & 3 & 5 & 2 & 9 \\ 
7 & 6 & 3 & 15 & 18 & 15 & 18 & 14 & 3 & 3 & 9 & 2%
\end{array}%
\right)
\end{equation}

Summing the diagonal elements gives $24$ ways. The eigenvalues of $T_{FBC}$
yield the same result: numerically, up to 15 significant figures, they are: $%
\lambda _{1}=4.561552812808830$, $\lambda _{2}=-3.414213562373095$, and $%
\lambda _{3}=$ $-3.414213562373093$. The last two values are probably the
same eigenvalue with multiplicity 2, since there is agreement in 14 digits.
One has further $\lambda _{4}=$ $1.999999999999998$, which probably is 2, $%
\lambda _{5}=$ $1.000000000000001$, $\lambda _{6}=\lambda
_{7}=1.000000000000000$. Finally $\lambda _{8}=\lambda
_{9}=-1.000000000000000$, $\lambda _{10}=-5.857864376269052\times 10^{-1}$, $%
\lambda _{11}=-5.857864376269045\times 10^{-1}$ (probably, again, multiple
eigenvalues), and $\lambda _{12}=$ $4.384471871911689\times 10^{-1}$.

From these results follow $\sum_{i=1}^{12}\lambda _{i}^{3}=24.\,\allowbreak
000\,000\,000\,\allowbreak 000\,035$, which says that we have 24 ways to
paint a world with 9 countries ($3\times 3$--lattice) with periodic boundary
condition in only one direction (within the numerical precision stated
above).

Now let us obtain $T_{PBC}$. This matrix can be obtained from $T_{FBC}$ by
excluding the columns and rows whose states have $C_{1}=C_{3}$, which
reduces the problem to a matrix of dimension $n_{\max }=6$.

\begin{equation}
T_{PBC}=\left( 
\begin{array}{ccccccc}
& \left\vert GYR\right\rangle & \left\vert RYG\right\rangle & \left\vert
GRY\right\rangle & \left\vert YRG\right\rangle & \left\vert YGR\right\rangle
& \left\vert RGY\right\rangle \\ 
\left\langle GYR\right\vert & 0 & 0 & 0 & 1 & 0 & 1 \\ 
\left\langle RYG\right\vert & 0 & 0 & 1 & 0 & 1 & 0 \\ 
\left\langle GRY\right\vert & 0 & 1 & 0 & 0 & 1 & 0 \\ 
\left\langle YRG\right\vert & 1 & 0 & 0 & 0 & 0 & 1 \\ 
\left\langle YGR\right\vert & 0 & 1 & 1 & 0 & 0 & 0 \\ 
\left\langle RGY\right\vert & 1 & 0 & 0 & 1 & 0 & 0%
\end{array}%
\right)
\end{equation}

First, we can test if we can obtain the result for periodic boundary
conditions in one direction. We need to calculated $Add(T_{PBC}^{2})$:

\begin{equation}
T_{PBC}^{2}=\left( 
\begin{array}{cccccc}
2 & 0 & 1 & 1 & 0 & 1 \\ 
0 & 2 & 1 & 0 & 1 & 0 \\ 
0 & 1 & 2 & 0 & 1 & 0 \\ 
1 & 0 & 0 & 2 & 0 & 1 \\ 
0 & 1 & 1 & 0 & 2 & 0 \\ 
1 & 0 & 0 & 1 & 0 & 2%
\end{array}%
\right)
\end{equation}%
\bigskip from which we obtain $Add(T_{PBC}^{2})=24$ exactly as we expect.
But, what if one wants the number of possibilities with periodic boundary
conditions in both directions? The eigenvalues of $T_{PBC}$ can in this case
be calculated exactly (numerically if needed) and their values are $\lambda
_{1}=\sqrt{2}+1,\lambda _{2}=1-\sqrt{2},\lambda _{3}=\sqrt{3},\lambda _{4}=-%
\sqrt{3},\lambda _{5}=\lambda _{6}=-1$. Therefore $\sum_{i=1}^{6}\lambda
_{i}^{3}=\left( \sqrt{2}+1\right) ^{3}-\left( \sqrt{2}-1\right)
^{3}-2=\allowbreak 12$.

In case one does not want to calculate eigenvalues, one may again take the
power $T_{PBC}^{3}$ and the trace:

\begin{equation}
T_{PBC}^{3}=\left( 
\begin{array}{cccccc}
2 & 1 & 1 & 3 & 1 & 4 \\ 
1 & 2 & 4 & 1 & 3 & 1 \\ 
1 & 4 & 2 & 1 & 4 & 5 \\ 
3 & 1 & 1 & 2 & 1 & 4 \\ 
1 & 3 & 4 & 1 & 2 & 1 \\ 
4 & 1 & 5 & 4 & 1 & 2%
\end{array}%
\right)
\end{equation}%
$\allowbreak$ which yields $Tr\,(T_{PBC}^{3})=12$, thus corroborating the
previous result. The problem, in this particular case, can be understood as
a $3\times\,3$ --\textit{sudoku}. We can count these 12 possibilities that
respect PBC in both directions. The simplest way to perform such counting by
hand is to fix the first color in the first country and to analyze all
possibilities for each choice. For each choice there are only 4
possibilities, since we can choose the first country in 3 different ways,
which yields a total of 12 as represented in Eq. \ref%
{Eq:twelve_possibilities}.

\begin{equation}
\begin{array}{ccccccc}
\begin{tabular}{|l|l|l|}
\hline
$\mathbf{R}$ & $G$ & $Y$ \\ \hline
$G$ & $Y$ & $R$ \\ \hline
$Y$ & $R$ & $G$ \\ \hline
\end{tabular}
&  & 
\begin{tabular}{|l|l|l|}
\hline
$\mathbf{R}$ & $G$ & $Y$ \\ \hline
$Y$ & $R$ & $G$ \\ \hline
$G$ & $Y$ & $R$ \\ \hline
\end{tabular}
&  & 
\begin{tabular}{|l|l|l|}
\hline
$\mathbf{R}$ & $Y$ & $G$ \\ \hline
$G$ & $R$ & $Y$ \\ \hline
$Y$ & $G$ & $R$ \\ \hline
\end{tabular}
&  & 
\begin{tabular}{|l|l|l|}
\hline
$\mathbf{R}$ & $Y$ & $G$ \\ \hline
$Y$ & $G$ & $R$ \\ \hline
$G$ & $R$ & $Y$ \\ \hline
\end{tabular}
\\ 
&  &  &  &  &  &  \\ 
\begin{tabular}{|l|l|l|}
\hline
$\mathbf{Y}$ & $R$ & $G$ \\ \hline
$R$ & $G$ & $Y$ \\ \hline
$G$ & $Y$ & $R$ \\ \hline
\end{tabular}
&  & 
\begin{tabular}{|l|l|l|}
\hline
$\mathbf{Y}$ & $R$ & $G$ \\ \hline
$G$ & $Y$ & $R$ \\ \hline
$R$ & $G$ & $Y$ \\ \hline
\end{tabular}
&  & 
\begin{tabular}{|l|l|l|}
\hline
$\mathbf{Y}$ & $G$ & $R$ \\ \hline
$R$ & $Y$ & $G$ \\ \hline
$G$ & $R$ & $Y$ \\ \hline
\end{tabular}
&  & 
\begin{tabular}{|l|l|l|}
\hline
$\mathbf{Y}$ & $G$ & $R$ \\ \hline
$G$ & $R$ & $Y$ \\ \hline
$R$ & $Y$ & $G$ \\ \hline
\end{tabular}
\\ 
&  &  &  &  &  &  \\ 
\begin{tabular}{|l|l|l|}
\hline
$\mathbf{G}$ & $Y$ & $R$ \\ \hline
$R$ & $G$ & $Y$ \\ \hline
$Y$ & $R$ & $G$ \\ \hline
\end{tabular}
&  & 
\begin{tabular}{|l|l|l|}
\hline
$\mathbf{G}$ & $Y$ & $R$ \\ \hline
$Y$ & $R$ & $G$ \\ \hline
$R$ & $G$ & $Y$ \\ \hline
\end{tabular}
&  & 
\begin{tabular}{|l|l|l|}
\hline
$\mathbf{G}$ & $R$ & $Y$ \\ \hline
$Y$ & $G$ & $R$ \\ \hline
$R$ & $Y$ & $G$ \\ \hline
\end{tabular}
&  & 
\begin{tabular}{|l|l|l|}
\hline
$\mathbf{G}$ & $R$ & $Y$ \\ \hline
$R$ & $Y$ & $G$ \\ \hline
$Y$ & $G$ & $R$ \\ \hline
\end{tabular}%
\end{array}
\label{Eq:twelve_possibilities}
\end{equation}

Since we analyzed the simplest cases by making explicit the matrices and
showing how the method works, we are prepared to numerically study the
problem and perform an extrapolation $N\rightarrow \infty $.

\section{Numerical results and finite-size effects via an efficient
computational method.}

\begin{figure}[tbp]
\begin{center}
\includegraphics[width=1.0\columnwidth]{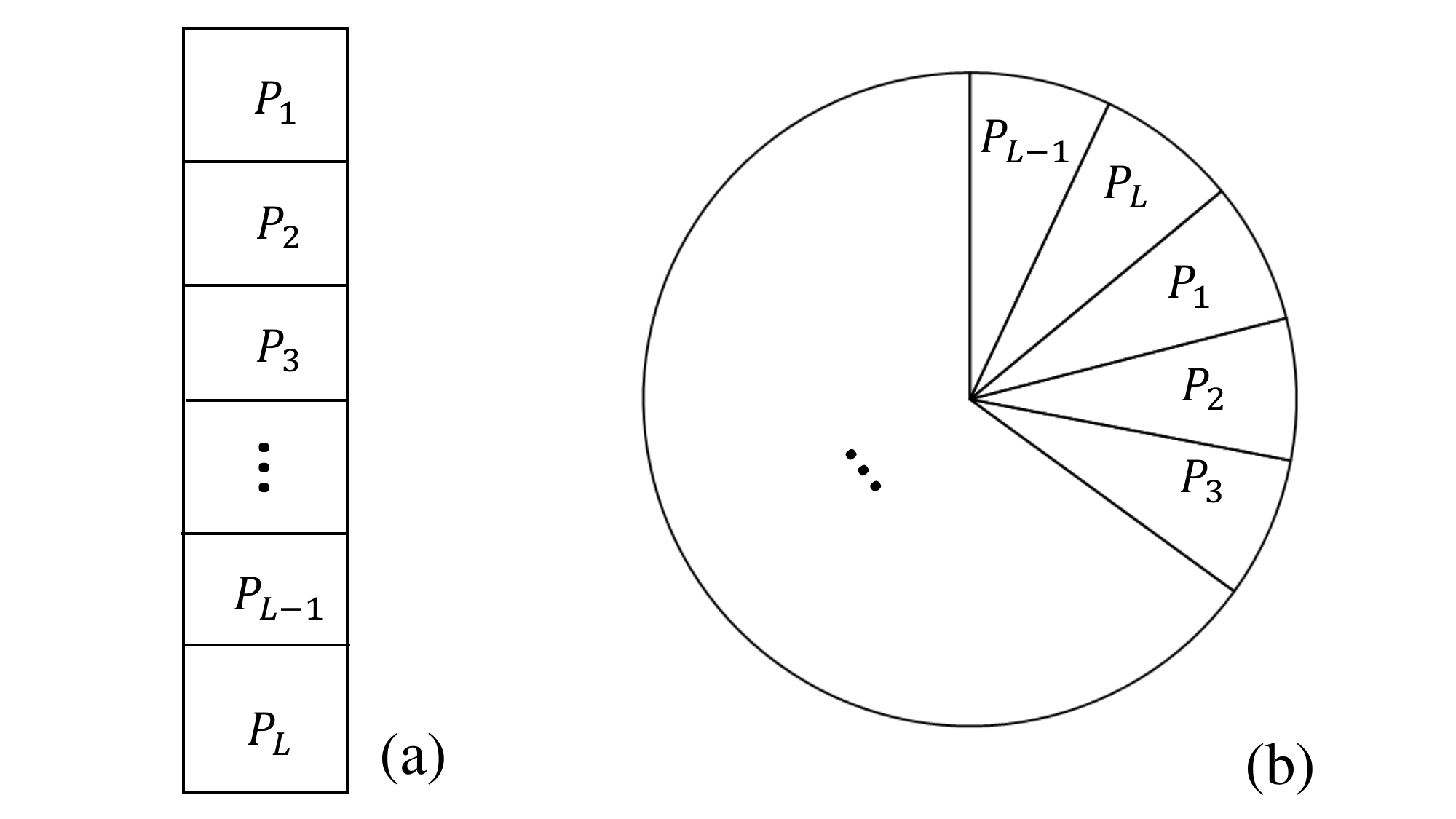}
\end{center}
\caption{Determination of the number of allowed states ($n_{\max
} $) for obtaining $T_{FBC}$ and $T_{PBC}$. }
\label{Fig_nmax}
\end{figure}

Our task begins by building the matrices $T_{FBC}$ and $T_{PBC}$, and find $%
n_{\max }$ (the number of states) as a function of $L$. But as we have seen
previously, this is not a difficult task: for example $n_{\max }$ for FBC is
exactly the number of ways of painting a unidimensional world with $L$
countries without periodic boundary conditions with three colors. This is
equivalent to paint a strip since $P_{1}$ is not a neighbor of $P_{L}$ (Fig. %
\ref{Fig_nmax}-a). On the other hand, for PBC $P_{1}$ is a neighbor of $%
P_{L} $, and this amounts to painting the circular sectors of a ''pizza'' as shown in Fig. \ref%
{Fig_nmax}-b.

In the first case we have 3 possibilities for $P_{1}$, but since $P_{1}$
cannot have the same color as $P_{2}$, there are 2 possibilities for the
latter. $P_{2}$ cannot share a color with $P_{3}$, which gives also 2
possibilities for $P_{3}$ and so on. With $L$ countries we have $n_{\max
}^{FBC}(L)=3\times 2^{L-1}$. However the second situation is more
complicate. Let us denote this number by $n_{\max }^{PBC}(L)$. We can think
for example the case where $P_{1}$ and $P_{3}$ have the same color or
different ones. In the first case, we have a situation where two sectors
merge into one and the same sector. Thus, for each coloring of the
pizza with $L-2$ sectors composed by the sector originated from the
fusion of $P_{1}$ with $P_{3}$ and all the remaining $L-3$ sectors (except $%
P_{2}$), one has 2 ways to paint the sector $P_{2}$ which can have neither
the same color of $P_{1}$ nor that of $P_{3}$. On the other hand if $P_{1}$
necessarily has a color different from $P_{3}$, so there is only one way to
color $P_{2}$. In this case we have to paint the pizza with $L-1$
sectors (all except $P_{2}$). Both situations are independent and we can
write a simple recurrence relation: $n_{\max }^{PBC}(L)=2\times n_{\max
}^{PBC}(L-2)+n_{\max }^{PBC}(L-1)$. Its solution leads to $n_{\max
}^{PBC}(L)=2^{L}+2(-1)^{L}$. (This can be checked by direct substitution)

Our task now is to build $T_{FBC}$ and $T_{PBC}$. We know the number of all
states, and each state is built in the following way:%
\begin{equation}
\left\vert \varphi \right\rangle _{FBC}=\left\vert C_{1}C_{2}...C_{L}|\delta
_{C_{i},C_{i+1}}=0\text{, for all }i=1...L-1\right\rangle
\end{equation}%
or%
\begin{equation}
\left\vert \varphi \right\rangle _{PBC}=\left\vert C_{1}C_{2}...C_{L}|\delta
_{C_{i},C_{i+1}}=0\text{ , for all }i=1...L\text{,\ and }C_{L+1}=C_{1}\right%
\rangle .
\end{equation}

To each of the allowed states $\left\vert \varphi ^{(j)}\right\rangle $,\ $%
j=1,...,n_{\max }$ a number $n(j)=\sum_{l=0}^{L-1}C_{l}^{(j)}4^{l}$ is
associated, with $C_{i}^{(j)}=0,\,1,\,2$ where we made the association $%
Y\rightarrow 0$, $G\rightarrow 1$, and $R\rightarrow 2$. So to determine the
matrix $T(j,j^{\prime })=\left\langle \varphi ^{(j)}|\varphi ^{(j^{\prime
})}\right\rangle $, we can use two interesting operators: the first one is
the \textit{exclusive-OR} operator ( IEOR($i_{1},i_{2}$) in Fortran) which
returns the bitwise Boolean exclusive-OR of integer $i_{1}$ and $i_{2}$,
that is if bits are the same, the result is 0. If not, it returns the result
1. For example if $i_{1}=(1001)_{2}$ and $i_{1}=(1111)_{2}$, one has IEOR($%
i_{1},i_{2}$) $=(0110)_{2}$. The second is an operator that returns a
logical true if the bit at $p$ in $q$ is set, and the counting of the bits
starts at 0. In FORTRAN, the operator has a syntax: BTEST($q,p$). For
example the IEOR($n(j),n(j^{\prime })$) makes the exclusive OR operation
\textquotedblleft bit by bit\textquotedblright\ of the two binaries $n(j)$
and $n(j^{\prime })$.

The result gives a binary sequence of $2L$ bits. Taking bits by pairs, i.e.,
the first and the second, third and the fourth, and so on, if one or more of
these pairs is 00, it implies that some neighboring countries share the same
color. This can be performed by executing the operation BTEST(IEOR($%
n(j),n(j^{\prime })$)$,k$) to check if bit $k$ is 0 or 1. Thus we can write
that:

\begin{equation}
T(j,j^{\prime })=\left\{ 
\begin{array}{ll}
1 & \text{if\ }\prod\limits_{\substack{ k=1  \\ k\text{ odd}}}^{2L-1}\left[ 
\text{BTEST}(\text{IEOR}(n(j),n(j^{\prime })),k)+\text{BTEST}(\text{IEOR}%
(n(j),n(j^{\prime })),k+1)\right] \neq 0 \\ 
&  \\ 
0 & \text{otherwise}%
\end{array}%
\right.  \label{Eq:T}
\end{equation}

Let us consider a particular case with $L=3$. For example $%
\left\vert \varphi ^{(j)}\right\rangle =$ $\left\vert
C_{1}^{(j)}C_{2}^{(j)}C_{3}^{(j)}\right\rangle =\left\vert 101\right\rangle
=1\times 4^{0}+0\times 4^{1}+1\times 4^{2}=17=1\times 2^{0}+0\times
2^{1}+0\times 2^{2}+0\times 2^{3}+1\times 2^{4}+0\times 2^{5}=(100010)_{2}$
and $\left\vert \varphi ^{(j^{\prime })}\right\rangle =$ $\left\vert
C_{1}^{(j^{\prime })}C_{2}^{(j^{\prime })}C_{3}^{(j^{\prime })}\right\rangle
=\left\vert 120\right\rangle =1\times 4^{0}+2\times 4^{1}+0\times
4^{2}=9=1\times 2^{0}+0\times 2^{1}+0\times 2^{2}+1\times 2^{3}+0\times
2^{4}+0\times 2^{5}=(100100)_{2}$. In this binary representation we can
consider that 00 corresponds to $Y$, 01 corresponds to $G$, and finally 10
corresponds to $R$. So the result IEOR($n(j),n(j^{\prime })$)$=$IEOR($17,9$)
can explicitly calculated as:%
\begin{equation}
\begin{tabular}{|c|ll|ll|ll|}
\hline\hline
\textbf{17} & 1 & 0 & 0 & 0 & 1 & 0 \\ \hline
\textbf{9} & 1 & 0 & 0 & 1 & 0 & 0 \\ \hline\hline
IEOR(17,9) & 0 & 0 & 0 & 1 & 1 & 0 \\ \hline\hline
\end{tabular}%
\end{equation}%
and one has that BTEST(IEOR($n(j),n(j^{\prime })$)$,1$) $=0$ and BTEST(IEOR($%
n(j),n(j^{\prime })$)$,1$)$=0$ and thus $T(j,j^{\prime })=0$ according Eq. %
\ref{Eq:T}.

Since we understand how to computationally build the matrices, we can show
our main results. We performed numerical experiments computing values using
both methods for a double-check: by directly calculating and then by
computing the eigenvalues.

\begin{table}[tbp] \centering%
\begin{tabular}{cccccc}
\hline\hline
$L$ & $n_{\max }^{(FBC)}$ & $Add(T_{FBC}^{L-1})$ & $Tr(T_{FBC}^{L})$ & $%
\sum_{i=1}^{n_{\max }}(\lambda _{FBC}^{(i)})^{L}$ & $\lambda _{FBC}^{(\max
)} $ \\ \hline\hline
\multicolumn{1}{l}{$2$} & \multicolumn{1}{r}{$6$} & \multicolumn{1}{r}{$18$}
& \multicolumn{1}{r}{$18$} & \multicolumn{1}{r}{$18$} & \multicolumn{1}{r}{$%
3 $} \\ 
\multicolumn{1}{l}{$3$} & \multicolumn{1}{r}{$12$} & \multicolumn{1}{r}{$246$%
} & \multicolumn{1}{r}{$24$} & \multicolumn{1}{r}{$24.000000000000040$} & 
\multicolumn{1}{r}{$4.561552812808830$} \\ 
\multicolumn{1}{l}{$4$} & \multicolumn{1}{r}{$24$} & \multicolumn{1}{r}{$%
7812 $} & \multicolumn{1}{r}{$4626$} & \multicolumn{1}{r}{$%
4625.999999999995000$} & \multicolumn{1}{r}{$6.971960768397091$} \\ 
\multicolumn{1}{l}{$5$} & \multicolumn{1}{r}{$48$} & \multicolumn{1}{r}{$%
580986$} & \multicolumn{1}{r}{$38880$} & \multicolumn{1}{r}{$%
38879.999999999990000$} & \multicolumn{1}{r}{$10.682885121208430$} \\ 
\multicolumn{1}{l}{$6$} & \multicolumn{1}{r}{$96$} & \multicolumn{1}{r}{$%
1.01596896\cdot 10^{8}$} & \multicolumn{1}{r}{$3.7284186\cdot 10^{7}$} & 
\multicolumn{1}{r}{$3.728418600000060\cdot 10^{7}$} & \multicolumn{1}{r}{$%
16.392041198957880$} \\ 
\multicolumn{1}{l}{$7$} & \multicolumn{1}{r}{$192$} & \multicolumn{1}{r}{$%
4.1869995708\cdot 10^{10}$} & \multicolumn{1}{r}{$1.886476032\cdot 10^{9}$}
& \multicolumn{1}{r}{$1.886476032000011\cdot 10^{9}$} & \multicolumn{1}{r}{$%
25.174078531617520$} \\ 
\multicolumn{1}{l}{$8$} & \multicolumn{1}{r}{$384$} & \multicolumn{1}{r}{$%
4.0724629633188\cdot 10^{13}$} & \multicolumn{1}{r}{$9.527634436194\cdot
10^{12}$} & \multicolumn{1}{r}{$9.52763443619383\cdot 10^{12}$} & 
\multicolumn{1}{r}{$38.683160866531850$} \\ 
\multicolumn{1}{l}{$9$} & \multicolumn{1}{r}{$768$} & \multicolumn{1}{r}{$%
9.357497524902707\cdot 10^{16}$} & \multicolumn{1}{r}{$2.825260002442752%
\cdot 10^{15}$} & \multicolumn{1}{r}{$2.825260002442852\cdot 10^{15}$} & 
\multicolumn{1}{r}{$59.465107914794780$} \\ 
\multicolumn{1}{l}{$10$} & \multicolumn{1}{r}{$1536$} & \multicolumn{1}{r}{$%
5.082795214936645\cdot 10^{20}$} & \multicolumn{1}{r}{$7.704801938642910%
\cdot 10^{19}$} & \multicolumn{1}{r}{$7.704801938642726\cdot 10^{19}$} & 
\multicolumn{1}{r}{$91.437962270582820$} \\ \hline\hline
\end{tabular}%
\caption{Results from $T_{FBC}$}\label{Table:I}%
\end{table}%

Tables \ref{Table:I} and \ref{Table:II} show the results for $T_{FBC}$ and $%
T_{PBC}$ respectively. We can observe as expected that the fourth (or fifty)
column in Table \ref{Table:I} leads to the same result represented in the third column
of Table \ref{Table:II}. The fifth column in Table \ref{Table:I} is just used as a
cross-check of the fourth column, exactly as in Table \ref{Table:II}.

\begin{table}[tbp] \centering%
\begin{tabular}{cccccc}
\hline\hline
$L$ & $n_{\max }^{(PBC)}$ & $Add(T_{PBC}^{L-1})$ & $Tr(T_{PBC}^{L})$ & $%
\sum_{i=1}^{n_{\max }}(\lambda _{PBC}^{(i)})^{L}$ & $\lambda _{PBC}^{(\max
)} $ \\ \hline\hline
\multicolumn{1}{l}{$2$} & \multicolumn{1}{r}{$6$} & \multicolumn{1}{r}{$18$}
& \multicolumn{1}{r}{$18$} & \multicolumn{1}{r}{$18$} & \multicolumn{1}{r}{$%
3 $} \\ 
\multicolumn{1}{l}{$3$} & \multicolumn{1}{r}{$6$} & \multicolumn{1}{r}{$24$}
& \multicolumn{1}{r}{$12$} & \multicolumn{1}{r}{$11.999999999999990$} & 
\multicolumn{1}{r}{$2.000000000000000$} \\ 
\multicolumn{1}{l}{$4$} & \multicolumn{1}{r}{$18$} & \multicolumn{1}{r}{$%
4626 $} & \multicolumn{1}{r}{$2970$} & \multicolumn{1}{r}{$%
2970.000000000001000$} & \multicolumn{1}{r}{$6.372281323269014$} \\ 
\multicolumn{1}{l}{$5$} & \multicolumn{1}{r}{$30$} & \multicolumn{1}{r}{$%
38880$} & \multicolumn{1}{r}{$7560$} & \multicolumn{1}{r}{$%
7559.999999999955000$} & \multicolumn{1}{r}{$5.999999999999998$} \\ 
\multicolumn{1}{l}{$6$} & \multicolumn{1}{r}{$66$} & \multicolumn{1}{r}{$%
3.7284186\cdot 10^{7}$} & \multicolumn{1}{r}{$1.64484\cdot 10^{7}$} & 
\multicolumn{1}{r}{$1.644839999999986\cdot 10^{7}$} & \multicolumn{1}{r}{$%
14.506431494048050$} \\ 
\multicolumn{1}{l}{$7$} & \multicolumn{1}{r}{$126$} & \multicolumn{1}{r}{$%
1.886476032\cdot 10^{9}$} & \multicolumn{1}{r}{$1.9900062\cdot 10^{8}$} & 
\multicolumn{1}{r}{$1.990006199999976\cdot 10^{8}$} & \multicolumn{1}{r}{$%
15.783341876392290$} \\ 
\multicolumn{1}{l}{$8$} & \multicolumn{1}{r}{$258$} & \multicolumn{1}{r}{$%
9.527634436194\cdot 10^{12}$} & \multicolumn{1}{r}{$2.901094068042\cdot
10^{12}$} & \multicolumn{1}{r}{$2.901094068041933\cdot 10^{12}$} & 
\multicolumn{1}{r}{$33.676786957721970$} \\ 
\multicolumn{1}{l}{$9$} & \multicolumn{1}{r}{$510$} & \multicolumn{1}{r}{$%
2.825260002442752\cdot 10^{15}$} & \multicolumn{1}{r}{$1.825277062836360%
\cdot 10^{14}$} & \multicolumn{1}{r}{$1.825277062836212\cdot 10^{14}$} & 
\multicolumn{1}{r}{$39.650566012033250$} \\ 
\multicolumn{1}{l}{$10$} & \multicolumn{1}{r}{$1026$} & \multicolumn{1}{r}{$%
7.704801938642287\cdot 10^{19}$} & \multicolumn{1}{r}{$1.617804974008648%
\cdot 10^{19}$} & \multicolumn{1}{r}{$1.617804974008630\cdot 10^{19}$} & 
\multicolumn{1}{r}{$78.818864591827550$} \\ \hline\hline
\end{tabular}%
\caption{Results from $T_{PBC}$}\label{Table:II}%
\end{table}%

But what does these results mean? To see that we can estimate from the
results obtained from the tables the quantity: 
\begin{equation}
W=(\frac{1}{3}\Omega )^{1/N}
\end{equation}%
which essentially is the exponential of the entropy per particle divided by
the Boltzmann constant. We will now perform an extrapolation $%
\lim_{N\rightarrow \infty }W=W_{\infty }$. In the next subsection, we will
present an ingenious method to do such an extrapolation by the use of
successive polynomial fits. Finally in subsection \ref{Subsec:BST_method} we
apply a more precise technique due to Bulirsch and Stoer \cite{BST1964} and
later applied to Statistical Mechanics \cite{BSTHenkel}.

\subsection{Polynomial Extrapolation}

\label{Subsec:Polynomial_Extrapolation}

\begin{figure}[tbp]
\begin{center}
\includegraphics[width=1.0\columnwidth]{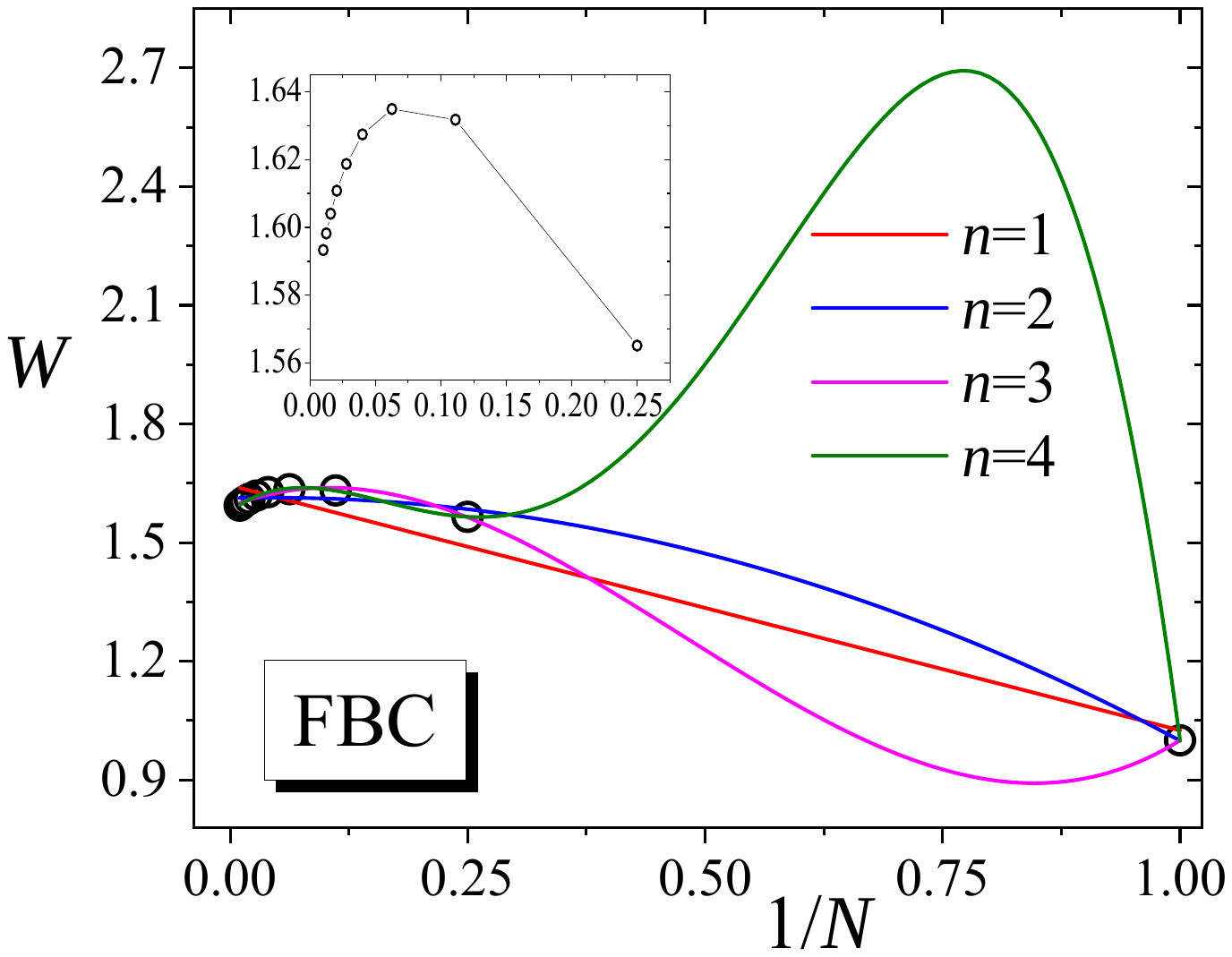}
\end{center}
\caption{$W$ $\times 1/N$ for FBC. We perform fits with polynomials of
degree 1, 2, 3, and 4. }
\label{Fig:FBC_extrapolation}
\end{figure}

First let us analyze the extrapolation $W$ $\times \ N^{-1}$ for $FBC$ which
can be seen in Fig. \ref{Fig:FBC_extrapolation}. Different polynomial fits
were performed to obtain an extrapolation $\lim_{N\rightarrow \infty
}W=W_{\infty }$. Line 2 of table \ref{Table:III} shows that the higher the
polynomial degree $n$ the better the extrapolated value $W_{\infty }$, since
the exact value for $W_{\infty }$ is $W_{Lieb}=1.5396007...$ (see ref. \cite%
{Lieb-II}). \ 
For FBC the results are not so good since the best result is $W_{\infty
}=1.572\pm 0.032$ (we are keeping the number of digits for trustworthiness).
It is important to observe that we are using

the simplest case for elaborating the plots, that is $L=1$ and $\Omega =3$.
In which case it is trivial to see that it does not depend on the boundary
conditions. Thus it is important to go beyond that, by studying the cases
1-PBC and 2-PBC.

\begin{table}[tbp] \centering%
\begin{tabular}{lllll}
\hline\hline
Boundary conditions & $n=1$ & $n=2$ & $n=3$ & $n=4$ \\ \hline\hline
FBC & 1.644(16) & 1.6122(81) & 1.5907(37) & 1.572(32) \\ 
1 - PBC (even branch) & 1.616(13) & 1.5868(46) & 1.5733(22) & 1.566(26) \\ 
2 - PBC (even branch) & 1.5360(16) & 1.539749(26) & 1.539675(22) & 
1.5395980(27) \\ 
1 - PBC (odd branch) & 1.466(54) & 1.5733(47) & 1.5608(18) & 1.555(15) \\ 
2 - PBC (odd branch) & 1.387(64) & 1.5143(81) & 1.5358(11) & 1.53947(13) \\ 
\hline\hline
\end{tabular}%
\caption{Values of $W$ for the different cases: FBC, 1-PBC, and 2-PBC. We also separate the results in even and 
odd branch to perform better extrapolations the corresponding uncertainties for $n$ = 1, 2, and 3 are due to 
uncertainty in the intercept after the polynomial extrapolation. The estimates for $n=4$ exactly appears 
with error bars since we used the difference between the extrapolated value an the exact value $(4/3)^(3/2)$.
It is important to mention that for this case the uncertainty in the intercept cannot be estimated since the 
number of points is exactly the number parameters in the interpolation} %
\label{Table:III}%
\end{table}%

\begin{figure}[tbp]
\begin{center}
\includegraphics[width=0.5\columnwidth]{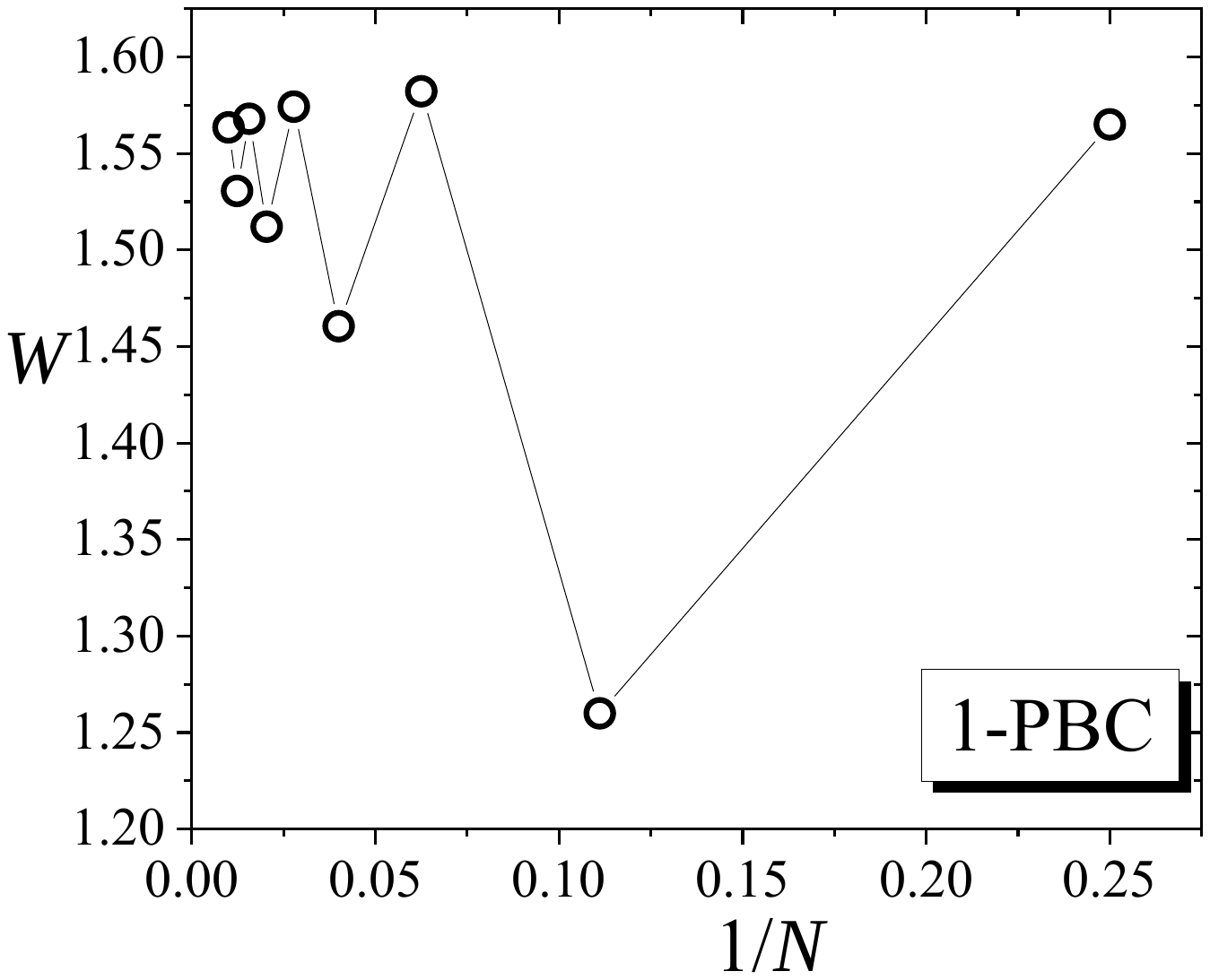}%
\includegraphics[width=0.5\columnwidth]{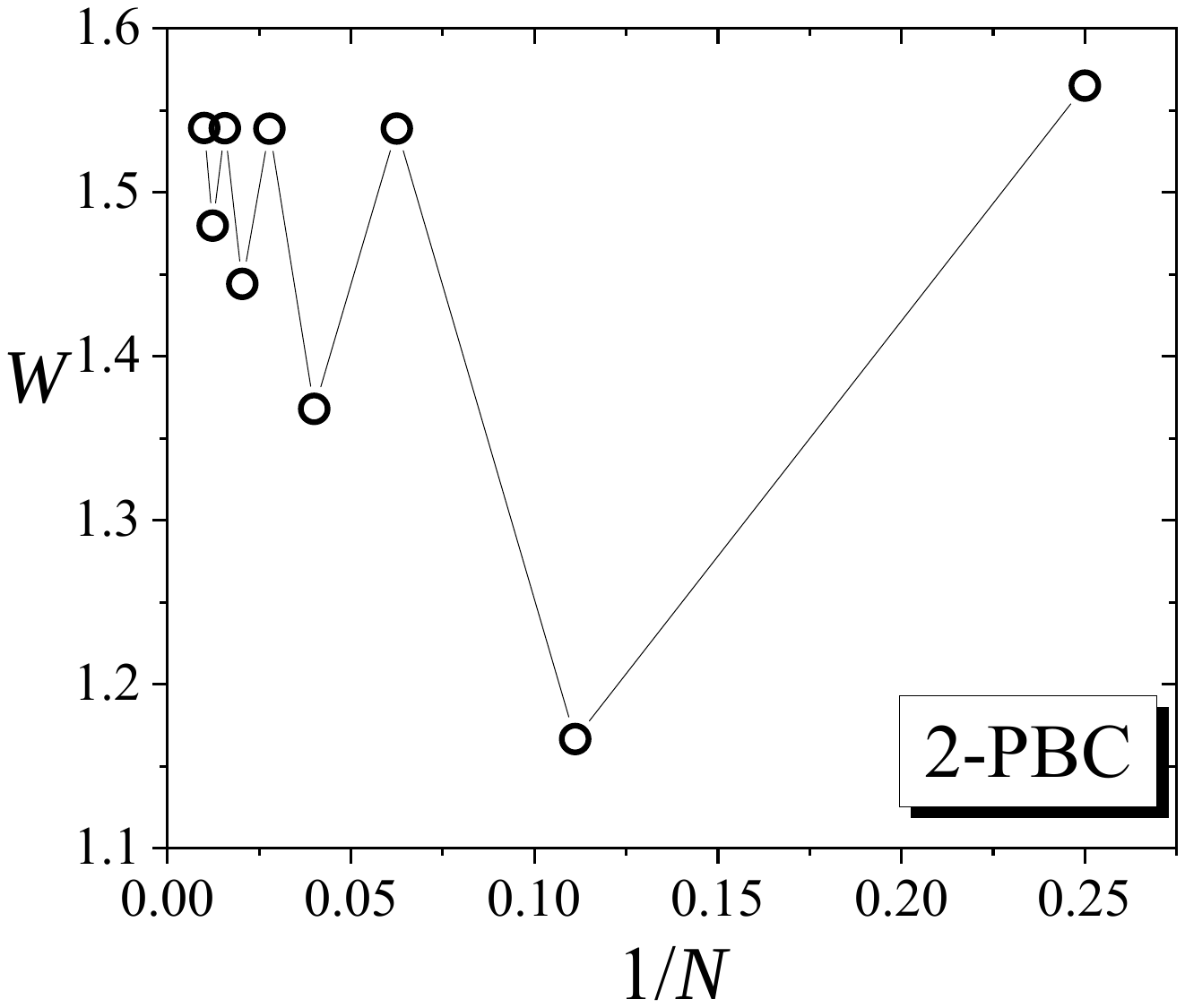} %
\includegraphics[width=0.8\columnwidth]{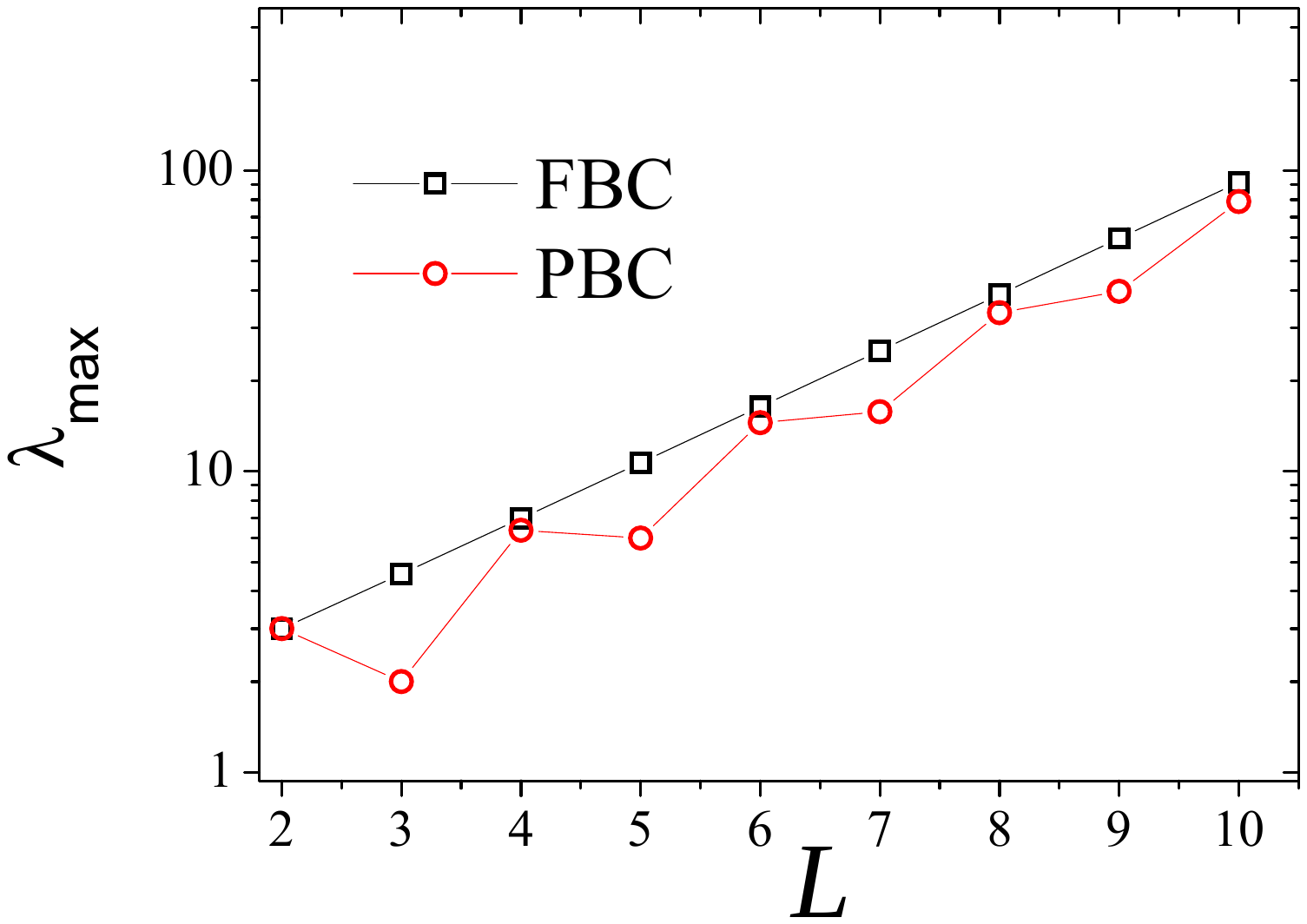}
\end{center}
\caption{$W$ $\times 1/N$ for 1-PBC (a) and 2-PBC (b). It is interesting to
observe that in the case of the PBC, an alternating convergence is observed.
In the plot (c), we observe that the highest eigenvalue of $T_{PBC}$ grows
up oscillating while $T_{FBC}$ does not present this oscillation between odd
and even $L$ values. }
\label{Fig:Alternate_PBC}
\end{figure}

We generated plots of $\ W$ $\times 1/N$ for 1-PBC, as depicted in Fig. \ref%
{Fig:Alternate_PBC} (a) and for 2-PBC, as depicted in Fig. \ref%
{Fig:Alternate_PBC} (b). It is interesting to observe that in the case of
PBC, an alternating convergence is observed. Just for the sake of
comparison, in Fig. \ref{Fig:Alternate_PBC} (c) we observe that the highest
eigenvalue of $T_{PBC}$ increases and oscillates while $T_{FBC}$ does not
present this oscillation between odd and even values of $L$, showing that
the highest eigenvalue seems to reflect the behavior observed in $W$ $\times
1/N$ in Figs. \ref{Fig:FBC_extrapolation} and \ref{Fig:Alternate_PBC} (a)
and (b).

\begin{figure}[tbp]
\begin{center}
\includegraphics[width=0.5\columnwidth]{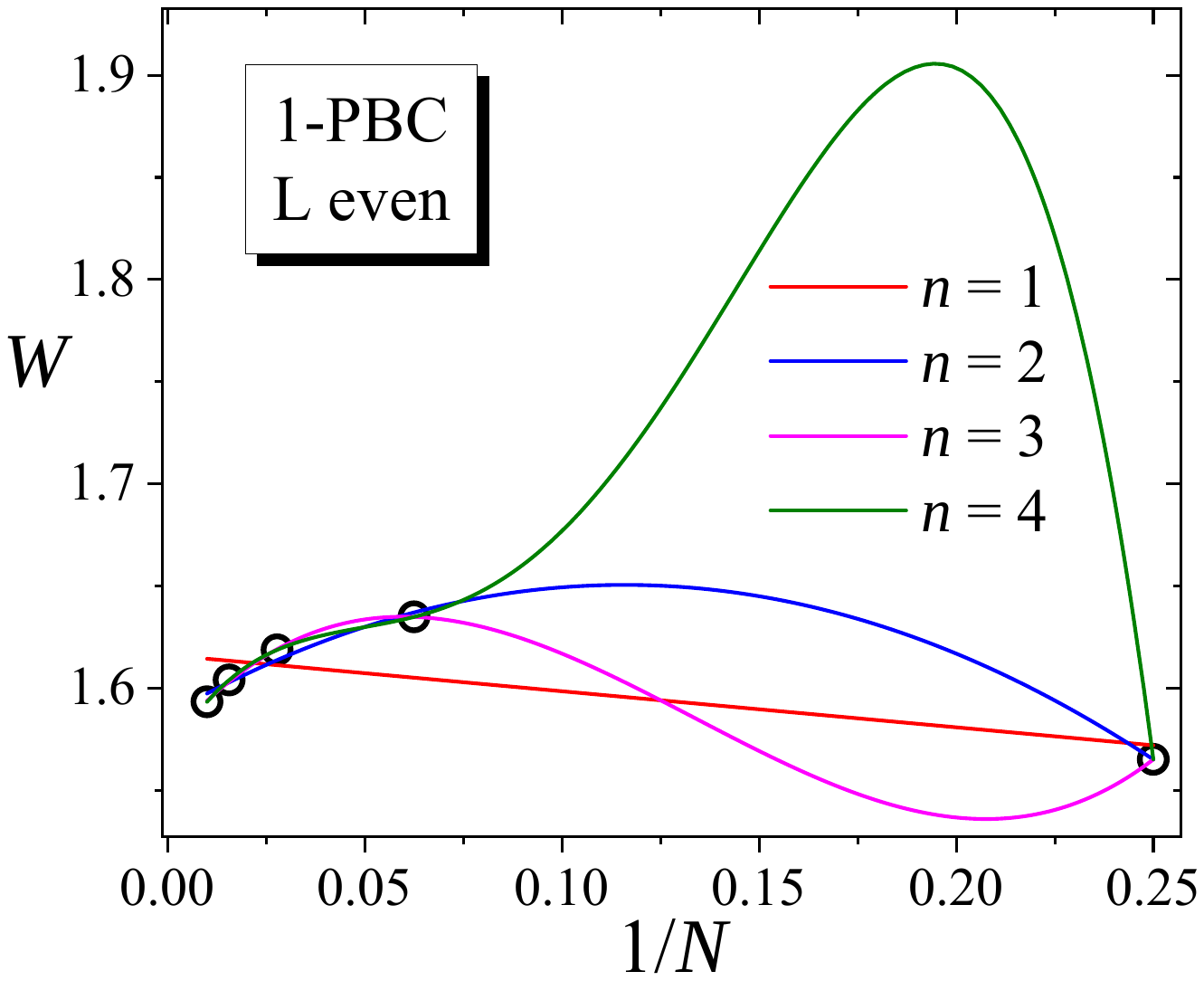}%
\includegraphics[width=0.5\columnwidth]{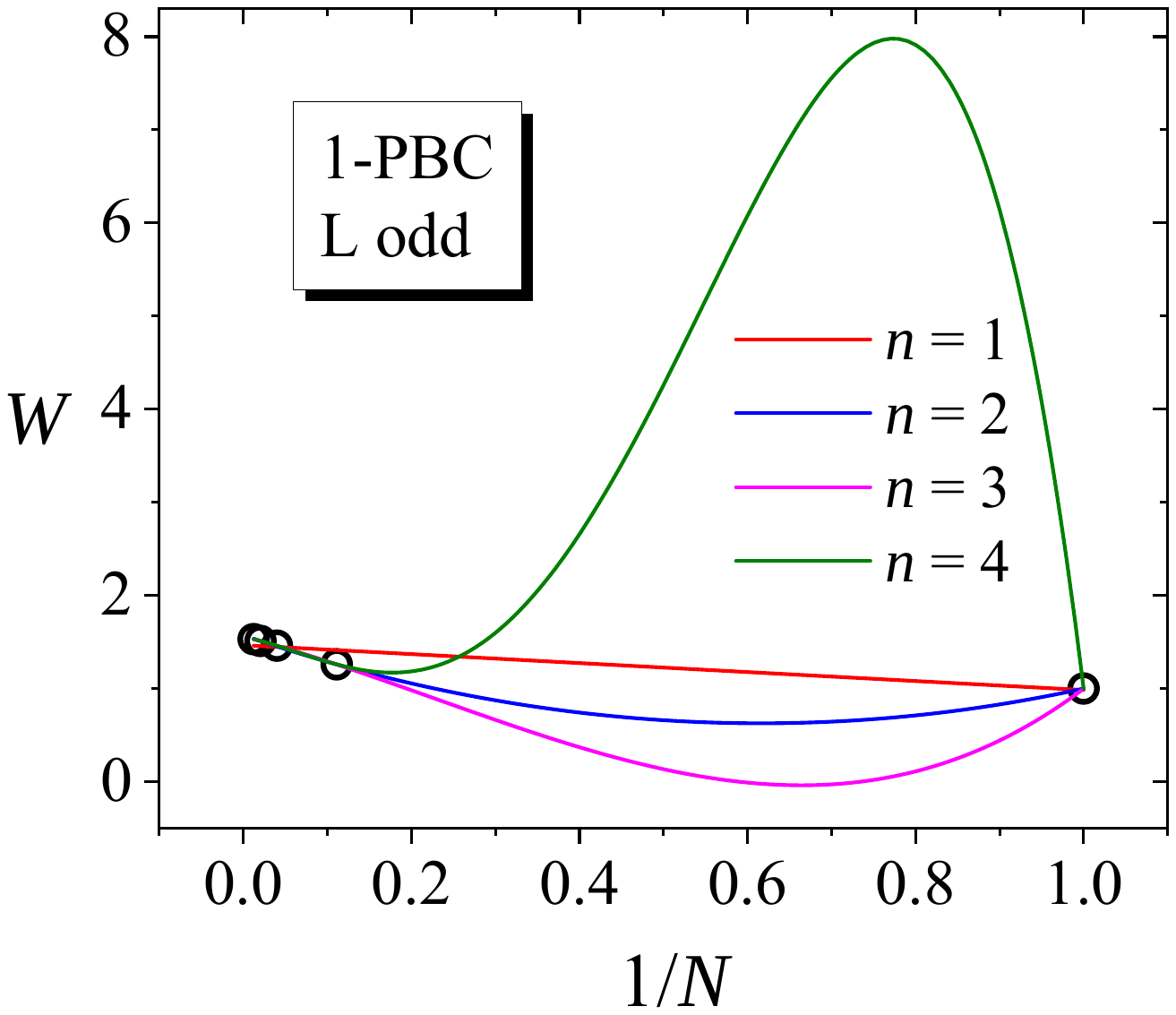} %
\includegraphics[width=0.5\columnwidth]{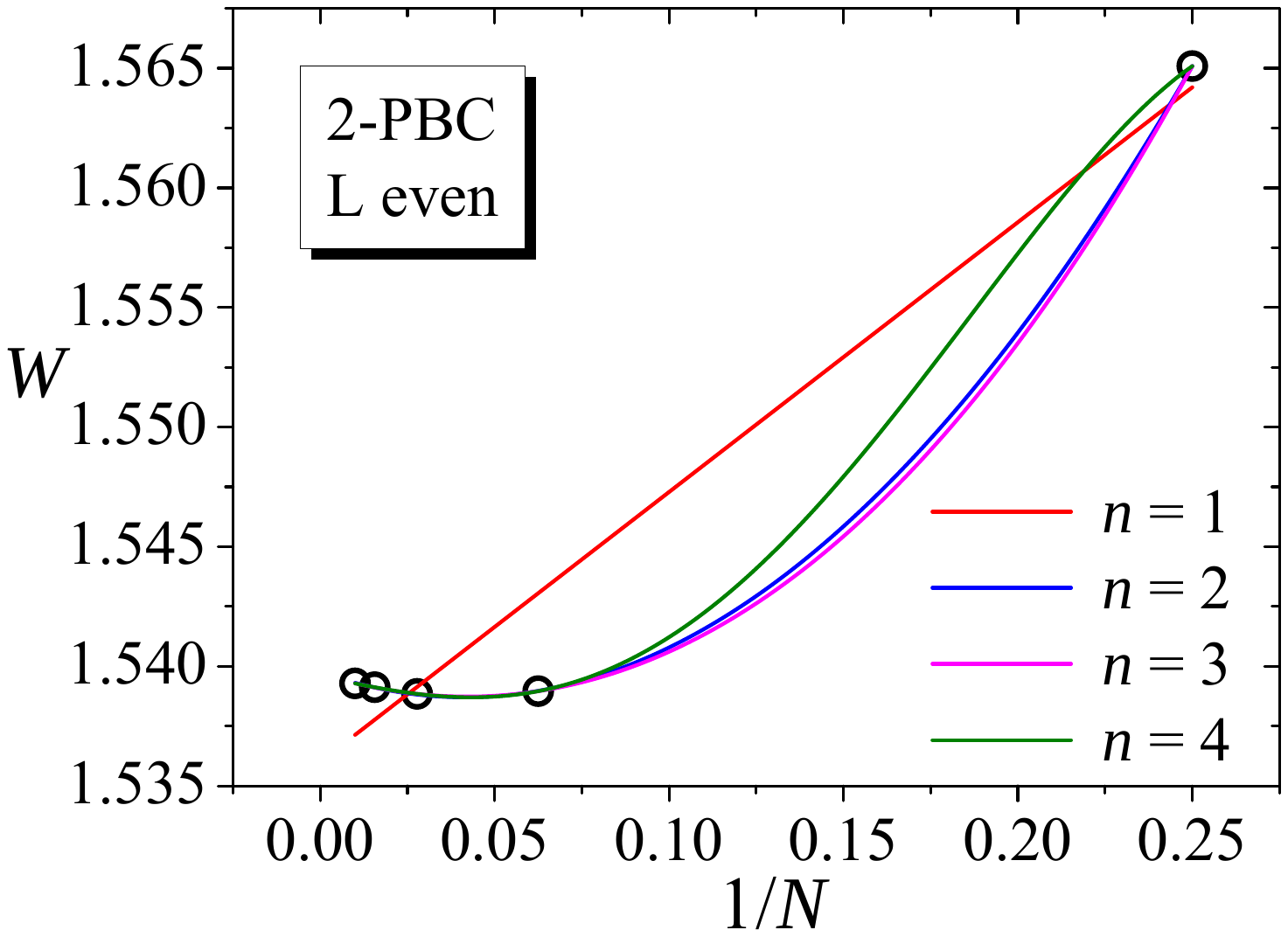}%
\includegraphics[width=0.5\columnwidth]{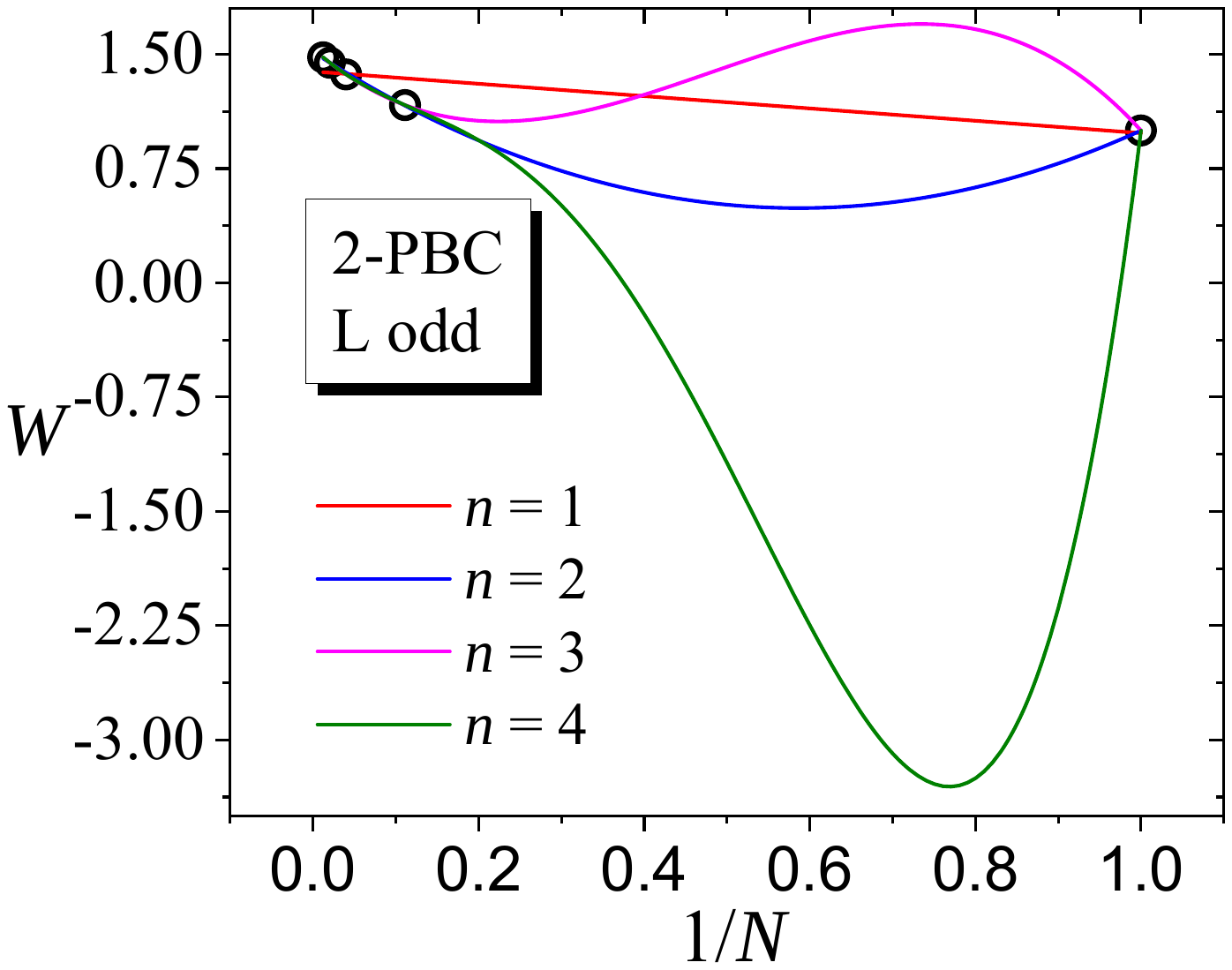}
\end{center}
\caption{Extrapolation of $W$ $\times 1/N$ for 1-PBC and 2-PBC. (a) Even
branch of 1-PBC, (b) Odd branch of 1-PBC, (c) Even branch of 2-PBC, and (d)
Odd branch of 2-PBC}
\label{Fig:Branches}
\end{figure}

Once we observed this oscillatory behavior in the convergence for PBC
between odd and even values of $L$, we separate the extrapolation in two
branches: odd and even. Figs. \ref{Fig:Branches} (a), (b), (c), and (d) show
the polynomial fits for each case considered/studied.

The rows in table \ref{Table:III} show the extrapolated value for each case.
We can observe that periodic boundary conditions in both directions
lead to a match with the exact result up to the fourth digit. Our results
with transfer-matrix leads to results better than we alternatively found in 
\cite{rdasilvaEJP}, however it is necessary to separate the extrapolation in
different branches. Such extrapolation can be obtained by using a method
really precise when compared with successive polynomial extrapolations with
higher degrees. In the following, we present the BST method by applying in
our model which gives really good estimates and mainly it does not require a
separation of the data in odd and even branches.\textbf{\ }

\subsection{Refined extrapolation: The Bulirsch-Stoer (BST) method}

\label{Subsec:BST_method}

By assuming that $W$\ is given by the expansion:%
\begin{equation*}
W=W_{\infty }+a_{1}h_{L}+a_{2}h_{L}^{2}+...
\end{equation*}%
where $h_{L}=\frac{1}{L^{2}}$, with $L=1,2,3,4...$, one get $W\rightarrow
W_{\infty }$ in the limit $L\rightarrow \infty $. The BST method considers a
sequence of extrapolants \cite{BSTHenkel}, which in our case can be given by:

\begin{equation}
\begin{array}{ccc}
W_{n,m} & = & W_{n+1,m-1}+\frac{\left( W_{n+1,m-1}-W_{n,m-1}\right) }{\left[
\left( \frac{h_{n+1}}{h_{n+m+1}}\right) ^{\omega }\left( 1-\frac{\left(
W_{n+1,m-1}-W_{n,m-1}\right) }{\left( W_{n+1,m-1}-W_{n,m-2}\right) }\right)
-1\right] } \\ 
&  &  \\ 
& = & W_{n+1,m-1}+\frac{\left( W_{n+1,m-1}-W_{n,m-1}\right) }{\left[ \left( 
\frac{_{n+m+1}}{n+1}\right) ^{2\omega }\left( 1-\frac{\left(
W_{n+1,m-1}-W_{n,m-1}\right) }{\left( W_{n+1,m-1}-W_{n,m-2}\right) }\right)
-1\right] }%
\end{array}
\label{Eq:recurrence}
\end{equation}%
where $\omega $\ is a free parameter.

Considering that one has the values of $W$\ for $L=1,2,...,L_{\max }$,
represented by $W_{0,0}$, $W_{1,0}$, $W_{2,0}$, ..., $W_{L_{\max }-1,0}$,
from Eq. \ref{Eq:recurrence} one obtains the subsequent values. However such
extrapolation are based on a binary tree, according to the table of
extrapolants:

\begin{equation}
\begin{array}{lllll}
W_{0,0} &  &  &  &  \\ 
& W_{0,1} &  &  &  \\ 
W_{1,0} &  & W_{0,2} &  &  \\ 
& W_{1,1} &  & \searrow &  \\ 
W_{2,0} & \vdots & \vdots & \cdots & W_{0,L_{\max }-1} \\ 
\vdots & W_{L_{\max }-3,1} &  & \nearrow &  \\ 
W_{L_{\max }-2,0} &  & W_{L_{\max }-3,2} &  &  \\ 
& W_{L_{\max }-2,1} &  &  &  \\ 
W_{L_{\max }-1,0} &  &  &  & 
\end{array}
\label{Eq:bynary_tree}
\end{equation}%
where with $W_{0,0}$\ and $W_{1,0}$, one obtains $W_{0,1}$, with $W_{1,0}$\
and $W_{2,0}$, one obtains $W_{1,1}$, and so successively. The same rule is
used to obtain the subsequent generations until one obtains $W_{0,L-1}$\
which is the best approximant to $W_{\infty }$, the root of the binary tree.
This is computationally performed by considering that for each $%
m=1,...,L_{\max }-1$, $n$ goes from $0$\ to $L-1-m$.

In order to determine the suitable $\omega _{opt}$\ we can take $\omega $
ranging from an $\omega _{\min }$\ until $\omega _{\max }$, with a
resolution $\Delta \omega $, to find the best $W_{0,L_{\max }-1}=W_{opt}$\
such that $\xi =\left\vert W_{0,L_{\max }-1}-W_{\infty }\right\vert $\ be
minimized since we know $W_{\infty }=\left( 4/3\right) ^{3/2}$.

\begin{figure}[tbp]
\begin{center}
\includegraphics[width=1.0\columnwidth]{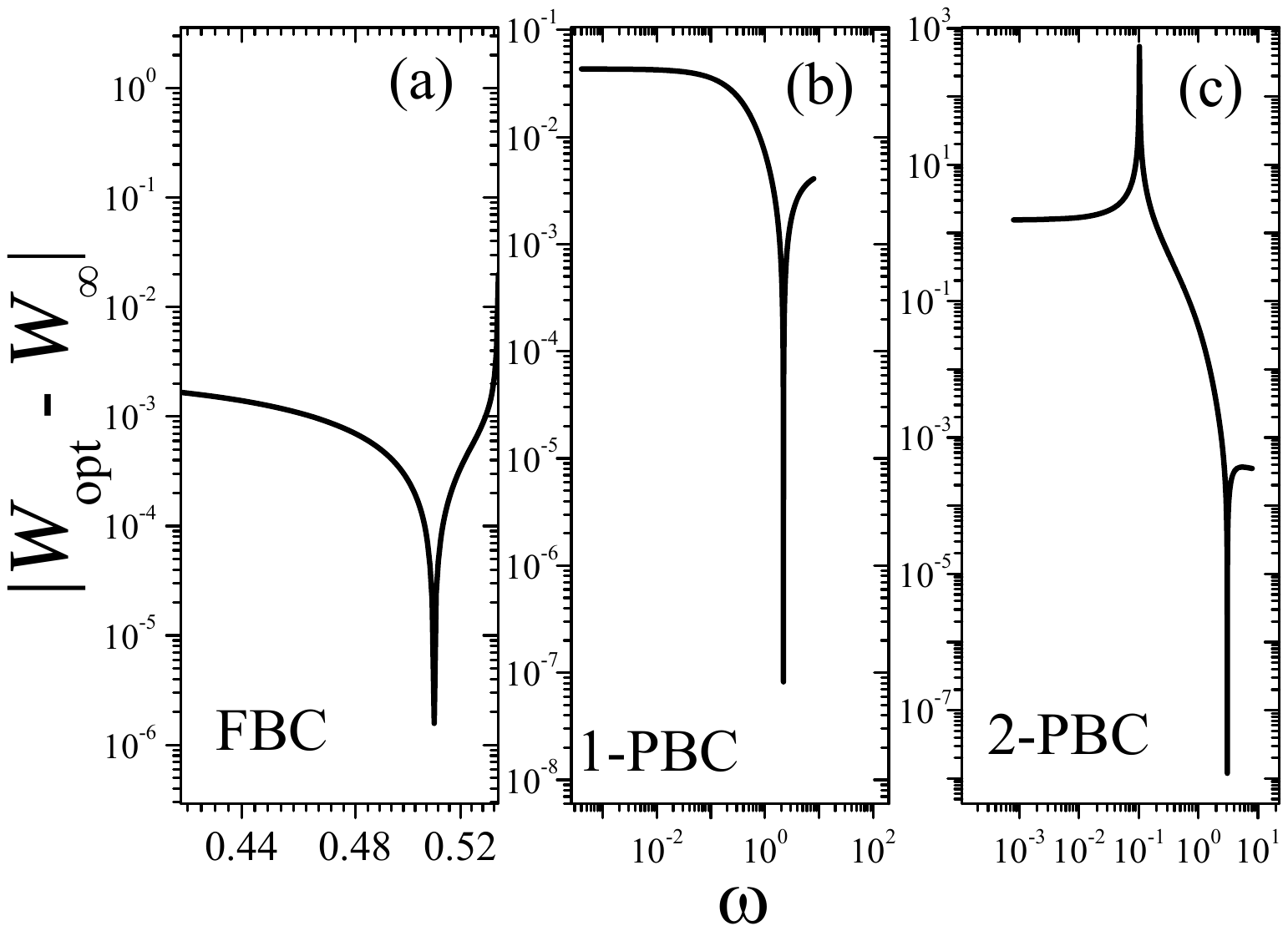}
\end{center}
\caption{$\protect\xi =\left\vert W_{opt}-W_{\infty }\right\vert $ as
function of $\protect\omega $ for FBC, 1-PBC, and 2-PBC }
\label{Fig:error_versus_omega}
\end{figure}
\ 

Fig. \ref{Fig:error_versus_omega} shows a plot of $\xi $\ versus $\omega $\
considering $\omega _{\min }=0$\ and $\omega _{\max }=8$, with $\Delta
\omega =8.10^{-4}$. For FBC (Fig. \ref{Fig:error_versus_omega} (a) ) and
2-PBC (Fig. \ref{Fig:error_versus_omega} (c) ) we found respectively $\omega
_{opt}=0.5096$\ and $\omega _{opt}=3.0664$, which yields $%
W_{opt}($FBC$)=1.539602\pm \allowbreak 0.000002$\ and $W_{opt}($2-PBC$%
)=1.53960073\pm 0.00000001$ respectively. By illustration, it is interesting to show the
sequence of iterations in the bynary tree as represented by Eq. \ref%
{Eq:bynary_tree}. For FBC one has the BST approximants represented in a
binary tree in Table \ref{Table:BSTFBC}.

\begin{table}[tbp] \centering%
\scriptsize
\begin{tabular}{lllllllllll}

\hline\hline
$W_{n,m}$ & $m=0$ & $m=1$ & $m=2$ & $m=3$ & $m=4$ & $m=5$ & $m=6$ & $m=7$ & $%
m=8$ & $m=9$ \\ \hline
&  &  &  &  &  &  &  &  &  &  \\ 
& $1.000000000$ &  &  &  &  &  &  &  &  &  \\ 
&  & $3.480583082$ &  &  &  &  &  &  &  &  \\ 
& $1.565084577$ &  & $1.715219692$ &  &  &  &  &  &  &  \\ 
&  & $1.779807114$ &  & $1.550250173$ &  &  &  &  &  &  \\ 
& $1.631721258$ &  & $1.638260759$ &  & $1.537231234$ &  &  &  &  &  \\ 
&  & $1.644097339$ &  & $1.540048426$ &  & $1.538396031$ &  &  &  &  \\ 
& $1.634848714$ &  & $1.630430852$ &  & $1.538278167$ &  & $1.539911992$ & 
&  &  \\ 
&  & $1.598650546$ &  & $1.538863512$ &  & $1.538867985$ &  & $1.539577074$
&  &  \\ 
& $1.627353072$ &  & $1.740465533$ &  & $1.538869734$ &  & $1.539605491$ & 
& $1.539654610$ &  \\ 
&  & $1.577487753$ &  & $1.538867110$ &  & $1.538862496$ &  & $1.539674735$
&  & $1.539602290$ \\ 
& $1.618676186$ &  & $1.449058272$ &  & $1.539171458$ &  & $1.539658715$ & 
& $1.539616720$ &  \\ 
&  & $1.565882904$ &  & $1.539021038$ &  & $1.542237835$ &  & $1.539517684$
&  &  \\ 
& $1.610780478$ &  & $1.514524921$ &  & $1.539345304$ &  & $1.539560599$ & 
&  &  \\ 
&  & $1.558840107$ &  & $1.539166123$ &  & $1.539683697$ &  &  &  &  \\ 
& $1.603980184$ &  & $1.527719176$ &  & $1.539434047$ &  &  &  &  &  \\ 
&  & $1.554254763$ &  & $1.539273370$ &  &  &  &  &  &  \\ 
& $1.598196268$ &  & $1.532769420$ &  &  &  &  &  &  &  \\ 
&  & $1.551109328$ &  &  &  &  &  &  &  &  \\ 
& $1.593271613$ &  &  &  &  &  &  &  &  &  \\ \hline\hline
\end{tabular}%
\caption{BST approximants for FBC with $L=1,2,...,10$}\label{Table:BSTFBC}%
\end{table}%

Similarly for the 2-PBC case, the sequence of iterations is represented in
table \ref{Table:BST2PBC}.

\begin{table}[tbp] \centering%
%
%
\scriptsize
\begin{tabular}{lllllllllll}
\hline\hline
$W_{n,m}$ & $m=0$ & $m=1$ & $m=2$ & $m=3$ & $m=4$ & $m=5$ & $m=6$ & $m=7$ & $%
m=8$ & $m=9$ \\ \hline
&  &  &  &  &  &  &  &  &  &  \\ 
& $1.000000000$ &  &  &  &  &  &  &  &  &  \\ 
&  & $1.577975700$ &  &  &  &  &  &  &  &  \\ 
& $1.565084580$ &  & $1.156646890$ &  &  &  &  &  &  &  \\ 
&  & $1.140183112$ &  & $1.5448184389$ &  &  &  &  &  &  \\ 
& $1.166529040$ &  & $1.542890160$ &  & $1.343729852$ &  &  &  &  &  \\ 
&  & $1.647705037$ &  & $1.3426505965$ &  & $1.540748058$ &  &  &  &  \\ 
& $1.538959531$ &  & $1.344749629$ &  & $1.540138974$ &  & $1.424988740$ & 
&  &  \\ 
&  & $1.317901102$ &  & $1.5646689105$ &  & $1.424844904$ &  & $1.539855817$
&  &  \\ 
& $1.367905267$ &  & $1.538874475$ &  & $1.425146964$ &  & $1.539590605$ & 
& $1.465328620$ &  \\ 
&  & $1.638264392$ &  & $1.4216357309$ &  & $1.549157927$ &  & $1.465299279$
&  & $1.539600730$ \\ 
& $1.538843701$ &  & $1.429591658$ &  & $1.539035457$ &  & $1.465363450$ & 
& $1.539462458$ &  \\ 
&  & $1.389882824$ &  & $1.5681790646$ &  & $1.464633278$ &  & $1.544169675$
&  &  \\ 
& $1.444199558$ &  & $1.539029999$ &  & $1.466430923$ &  & $1.539171586$ & 
&  &  \\ 
&  & $1.623256319$ &  & $1.4588169435$ &  & $1.552046417$ &  &  &  &  \\ 
& $1.539120096$ &  & $1.470888365$ &  & $1.539170223$ &  &  &  &  &  \\ 
&  & $1.427676251$ &  & $1.5392230356$ &  &  &  &  &  &  \\ 
& $1.479705887$ &  & $1.539223036$ &  &  &  &  &  &  &  \\ 
&  & $1.610686315$ &  &  &  &  &  &  &  &  \\ 
& $1.539281498$ &  &  &  &  &  &  &  &  &  \\ \hline\hline
\end{tabular}%
\caption{BST approximants for 2-PBC with $L=1,2,...,10$}\label{Table:BST2PBC}%
\end{table}%

We considered the 1-PBC as a separate case, since its convergence was more
delicate and thus $L=10$\ was not enough to obtain good estimates via BST
method. Based on this point, we extended our initial values considering $%
L=15 $. In this case we really obtained good results as shown in Fig.\ref%
{Fig:error_versus_omega} (b), which gives $\omega _{opt}($1-PBC$)=2.1996$,
resulting in $W_{opt}($1-PBC$)=1.53960063\pm 0.00000008$\ which supplies a
much better estimate when compared with the linear fit performed in \cite%
{rdasilvaEJP}.

We summarized our main results from BST in Table \ref{Table:Final_results}.

\begin{table}[tbp] \centering%
\begin{tabular}{lll}
\hline\hline
Lattice & $\omega _{opt}$ & $W_{opt}$ \\ \hline
&  &  \\ 
FBC & $0.5096$\textbf{\ } & $1.539602(2)$ \\ 
&  &  \\ 
1-PBC & $2.1996$ & $1.53960063(8)$ \\ 
&  &  \\ 
2-PBC & $3.0664$ & $1.53960073(1)$ \\ \hline\hline
\end{tabular}%
\caption{Final results from BST approximants}\label{Table:Final_results}%
\end{table}%

\subsection{Brief comments about the computational efforts}

Our method considers the use of explicit matrices. Basically, the main point
of this paper is the extension for periodic boundary conditions in both
directions which necessarily demands at least a complexity $O(n_{\max }^{3})=$ $O(8^{L})\ 
$ operations in any of the alternatives: taking powers of $n_{\max }\times n_{\max }$
matrices or simply calculating the set of eigenvalues of a single matrix $%
n_{\max }\times n_{\max }$. Operations to prepare the states and build the
matrices are relatively faster: $O(4^{L})$. In this case, the complexity is
better represented by $O(8^{L})$. It is computationally \textquotedblleft
expensive\textquotedblright\ since the exponent is linear in $L$, but the
exponential dependence can be a nuisance for larger lattices. \ 

Our intention was to show the strength of the method even when not
considering large lattices. For example, a simple and interesting
extrapolation method as BST works since our task is to deal with $8^{L}$ and
not simply $L$. For example, using a processor Intel(R) Core(TM) i7-8565U
CPU @ 1.80GHz-1.99 GHz with IMSL Numerical Libraries, for $L=10$ one needs a
few seconds while for $L=12$ something around 2 minutes. This is fine for
such sizes but one has to be careful: if one considers a not so large
lattice size $L=15$, one estimates a processing time around $t\approx \frac{%
8^{15}}{8^{12}}\times 2\ $min$\ =2^{10}$min, something near $t\approx 17$
hours which is still feasible but it starts to become \textquotedblleft
indigestible\textquotedblright 

Without periodic boundary conditions in both directions, one can use an
implicit method where one does not need to explicitly build the matrix (see 
\cite{rdasilvaEJP}). This method is highly efficient but it only works (to
the best of our knowledge) for free and periodic boundary conditions in one
direction.

\section{Conclusions}

This work extends the numerical transfer matrix method proposed by Creswick 
\cite{Creswick} to contemplate the case of toroidal boundary conditions. The
problem we deal with is the three-color one, which in turn is equivalent to
the six-vertex model proposed by Pauling to explain the residual entropy of
ice at $T=0$. Our results (see table \ref{Table:Final_results}) agree very
well with the exact result obtained by Lieb \cite{Lieb-I,Lieb-II} for square
lattices $W_{Lieb}=(4/3)^{3/2}\approx 1.5396007$ by solving exactly the
six-vertex model, or by Baxter \cite{Baxter} when mapping the problem in a
hard-square lattice gas, or even by Biggs \cite{Biggs} when obtaining the
chromatic polynomial for a finite toroidal square lattice graph exactly for
three colors. No wonder all results are obtained in the thermodynamic limit
and 3 colors is also very important since the number of ways to properly
paint a square lattice graph with $x$ colors, $\phi (x)$, satisfies the
inequality in the thermodynamic limit \cite{Biggs}:

\begin{equation*}
\frac{1}{2}(x-2+\sqrt{x^{2}-4x+8})\geq \phi (x)\geq \frac{x^{2}-3x+3}{x-1}
\end{equation*}%
and for $x=3$, the both sides of this inequality assume the same value: $%
\phi (3)=(4/3)^{3/2}=W_{Lieb}$. We differently worked with finite systems
and with a subsequent extrapolation which leads to this same value,
suggesting that our method could be extended for other models in statistical
mechanics, where the thermodynamic limit is not known.

Finally, we also show how to obtain the results for periodic boundary
conditions in only one direction from a matrix constructed with free
boundary conditions and we also show how to obtain toroidal boundary
conditions from a matrix previously constructed with periodic boundary
conditions in one direction, which shows the flexibility of our approach. It
is important to mention that our result for $W$\ is essentially the same for
all boundary conditions and the uncertainties in table \ref%
{Table:Final_results} show the numerical precision that such approaches may
have.

\section*{Acknowledgments}

R. da Silva thanks CNPq for financial support under grant numbers
311236/2018-9, and 424052/2018-0. The authors would like to thank the
anonymous referee for suggesting the use of the BST method to perform our
extrapolations. The authors also thank the Prof. P. Nightingale for kindly
sending us the reprints of his manuscript~\cite{Nightingale}.

\end{document}